\documentclass[Crown, sagev, doublespace, 11pt]{sagej}
\usepackage{moreverb,url, dsfont, soul, dsfont,amsmath,tikz, natbib}
\usepackage[colorlinks, bookmarksopen, bookmarksnumbered, citecolor=red, urlcolor=red]{hyperref}
\usepackage{moreverb, url, soul, subcaption, threeparttable, multicol, multirow, lscape, makecell, bm, bbm, soul}
\usepackage{enumitem}
\usepackage[title]{appendix}
\usepackage{mathtools}
\usepackage{geometry}
\usepackage{tabularx, ltablex, floatrow, inputenc}
\usepackage{xr}
\usepackage{xr-hyper}
\usepackage{hyperref}
\newcommand\BibTeX{{\rmfamily B\kern-.05em \textsc{i\kern-.025em b}\kern-.08em
		T\kern-.1667em\lower.7ex\hbox{E}\kern-.125emX}}

\def\volumeyear{2019}
\newcommand{\bigCI}{\mathrel{\text{\scalebox{1}{$\perp\mkern-10mu\perp$}}}}
\begin{document}
	
		\runninghead{Matsouaka et al.}
	\title{Variance estimation for the average treatment effects on the treated and on the controls.} 
	\author{Roland A. Matsouaka\affilnum{1,}\affilnum{2}, Yi Liu\affilnum{1}, and Yunji Zhou\affilnum{1}
	}
	
	\affiliation{\affilnum{1} Department of Biostatistics and Bioinformatics, Duke University, Durham, North Carolina\\
		\affilnum{2} Program for Comparative Effectiveness Methodology, Duke Clinical Research Institute,  Durham, North Carolina
	}
	\corrauth{Roland A. Matsouaka, Duke Clinical Research Institute,
		200  Morris St., Room 7826
		Durham, NC 27701}
	
	\email{roland.matsouaka@duke.edu}
	
	\begin{abstract}\small
		Common causal estimands include the average treatment effect (ATE), the average treatment effect of the treated (ATT), and the average treatment effect on the controls (ATC). Using augmented inverse probability weighting methods, parametric models are judiciously  leveraged to yield doubly robust estimators, i.e., estimators that are consistent when at least one the parametric models is correctly specified. Three sources of uncertainty are associated when we evaluate these estimators and their variances, i.e., when we estimate the treatment and outcome regression models as well as the desired treatment effect. 
		
		In this paper, we propose  methods to calculate the variance of the normalized, doubly robust ATT and ATC estimators and investigate their finite sample properties.  We consider both the asymptotic sandwich variance estimation, the standard bootstrap as well as two wild bootstrap methods. For the asymptotic approximations, we incorporate the aforementioned uncertainties via estimating equations. Moreover, unlike the standard bootstrap procedures, the proposed wild bootstrap methods use perturbations of the influence functions of the estimators through independently distributed random variables. We conduct an extensive simulation study where we vary the heterogeneity of the treatment effect as well as the proportion of participants assigned to the active treatment group. We illustrate the methods using an observational study of critical ill patients on the use of right heart catherization.
		
	\end{abstract}
	\maketitle	

	\section{Introduction}\label{sec:introduction}
	 Propensity score (PS) methods are widely  used in non-randomized studies to adjust for confounding and estimate treatment effects. The propensity score is the probability of receiving a specific treatment,  conditional on the participants' measured characteristics. 	
	 The interesting feature of propensity score methods is that they allow us to correct for covariate distributions imbalance between the treatment groups, leading to a randomized-clinical-trial-like situation, without looking at the outcome.\cite{hernan2016using,hernan2011great} 
	 
	 In practice, we often use a semiparametric method to assess the treatment effect: we estimate propensity scores by modeling the treatment allocation as a function of measured covariates and combine these scores to a non-parametric estimation of the treatment effect. If the propensity score model is correctly specified, then the distribution of the baseline covariates included in the propensity score model will be balanced. This allows unbiased estimation of treatment effect at each value of the propensity score. 
	 
	 There are several PS-based methods to estimate the treatment effect, including matching, propensity score stratification, and inverse probability weighting (IPW) by functions of the propensity scores.\cite{stuart2010matching, zhou2020propensity} In this paper, we are concerned with the use of the inverse probability weighting methods to estimate treatment effects in propensity score analysis. For IPW, this entails (1) estimating the propensity scores and (2) weighting each treatment and control participants, by functions of the propensity scores, and comparing the average weighted outcome variable between the treatment groups, based on specific estimand and the target population of interest. Besides the average treatment effect (ATE), other causal estimands of interest include the average treatment effect on the treated (ATT) individuals and the average treatment effect on the control (ATC) individuals. The latter two estimands are often of particular interest in policy evaluation(s)  and will be the subject of this paper. 
	 
	 Whether one chooses to use ATT or ATC depends primarily on their scientific question. The ATT  helps answer the question ``How would the outcome(s) differ, on average, in the treated patients, had they not received the treatment?''---in the context where it may be more informative to 
	 assess 	 the impact of the treatment on the participants who took it.\cite{greifer2021choosing,austin2021applying}  This question is often asked with a broader perspective in mind where the ATT is used to evaluate the effect of withholding the treatment from those in the population who otherwise receive it (i.e., resemble our study treated participants). This could be, for instance, the average effect of withholding (versus not withholding) the use of antidepressants during pregnancy on the risk of autism spectrum disorder in children, among women who are prescribed such medications. \cite{moodie2018doubly}   Additional examples can be found, across a number of scientific disciplines, where ATT was used as the target estimand. \cite{moodie2018doubly,reifeis2020variance} The  ATC, on the other hand, helps evaluate the impact of rolling out a treatment (resp. a policy) to those who are not currently receiving (resp. participating in) it. As an estimand, ATC is used to answer the question ``Should we expand this treatment---known to be effective for some patients---to those in the population who have not received it yet or otherwise won't receive it?'' \cite{greifer2021choosing} 
	  
	 While several papers have been published on how to (point) estimate ATT and ATC,\cite{moodie2018doubly} only few papers have proposed and studied the performance of the variance estimators of ATT and ATC---including recent papers by Moodie et al.\cite{moodie2018doubly} and Reifeis and Hudgens. \cite{reichardt1999justifying} Our paper extends those two papers by looking at the doubly robust estimators of ATT and ATC that also include outcome regression models.	 Doubly robust estimators are consistent for their targeted estimands when at least the propensity score (PS) model or the outcome regression (OR) models are correctly specified. In addition, these estimators are locally efficient  (i.e., have minimum variance) when PS and OR models are correctly specified.\cite{tsiatis2007semiparametric}  Beyond point estimation, we need to calculate appropriate measures of uncertainty estimates, such as variance or confidence interval, which are required to draw proper inference. Consequently, efficient estimation of the variance of these doubly robust estimators relies intrinsically on whether both the propensity and outcome regression models are correctly specified. 
	 Our objective, therefore, is to provide and investigate both the asymptotic  and the bootstrap variance estimators of the doubly robust estimators of the ATT and ATC and assess their performance. We specifically consider their normalized-weighted doubly robust estimators and demonstrate the importance of taking into account the uncertainties associated with the estimation of the propensity scores in deriving the variance estimates of these two estimands. {\color{black} In practice,  statisticians prefer to use sandwich variance estimation since it is based on asymptotic normal distribution, which often achieves robust estimates of the said variance. Standard bootstrap is often the go-to variance estimation method when a formula such as the asymptotic sandwich variance estimation is not readily available or applicable due to a small sample size.\cite{efron1994introduction,austin2014use} In the former case, for instance, we may not be able to estimate the variance accurately, especially if the postulated parametric models are unstable (or misspecified) or some treatment groups contain very few observations. In the latter case, the modus operandi of the standard bootstrap, which consists of resampling with repetition, may  exacerbates some issues we want to avoid. Resampling with repetition when the sample size is small can easily lead to unstable propensity score models and extreme weights. This can then introduce bias in point estimation and thus affect the variance estimation.\cite{zhou2020propensity} As an alternative variance estimation, we introduce the wild bootstrap methods that use directly the efficient influence function of the estimator, which is known to lead to efficient estimation of the variance.\cite{tsiatis2007semiparametric} Investigating the performance of these variance estimation methods under different scenarios allows us to understand which ones work better in what context(s).}
	 
	 We have organized the paper as follows. In Section \ref{sec:estimation}, we define the inverse probability weighting estimators for the ATT and ATC, along with the corresponding doubly robust estimators. The different large-sample variance estimators (based on asymptotic approximations and wild  bootstrap procedures) are introduced in Section \ref{sec:variance_estimation} for these four estimators. More importantly these estimators are robust to any heterocedastic (mis)specifications of the outcome mean-variance relationships. On one hand, the asymptotic variance estimation method exploits M-theory and related estimating equations to determine normal approximations to the distributions of the estimators and derive the related variances. Key mathematical derivations of these formulas, which are not readily available from other sources, are provided in details in the Supplemental Material \ref{sec:sandvariance_estimation}---for the case where the propensity scores and the outcome are modeled via using, respectively, logistic and linear regression models. The wild bootstrap method, on the other hand, leverages the characteristics of the efficient influence functions of the estimands and determines the corresponding variance estimations by perturbing these influence functions.   Then, in Section \ref{sec:simulations} we conduct and present simulation studies to compare the performance of the proposed variance estimators under finite sample sizes. We illustrate in Section \ref{sec:examples} the variance estimation methods in two data sets. Finally, we  conclude the paper in Section \ref{sec:conclusion}. 
	
	\section{Estimation framework}\label{sec:estimation}
	\subsection{Causal estimands}\label{sec:estimands}
	Let $Z$ denote  the indicator of treatment (or sex, race, or program participation), with $Z=1$ for treated and  $Z=0$ for control;    $Y$ a continuous outcome;  and ${X}=(X_0, X_1, \ldots, X_p)$ a matrix of baseline covariates, where the column vector $X_0=(1,\ldots, 1)'$.  The observed data $O=\{ (Z_i, X_i, Y_i): i=1,\dots, N \}$ are a sample of $N$ participants drawn independently from a large population of interest.    
	
	We adopt the potential outcome framework of Neyman-Rubin \cite{neyman1923applications, imbens2015causal}, i.e., each participant has two potential outcomes $Y(0)$ and $ Y(1)$,  where $Y(z)$ is the  outcome that would occur if, possibly contrary to fact, the individual were to receive treatment $Z=z.$ Potential outcomes are related to observed outcomes via 
	$Y=ZY(1)+(1-Z)Y(0),$ i.e., for each individual, the potential outcome  $Y(z)$  matches their observed outcome  $Y$ for the treatment  $Z=z$ they indeed received, by the consistency assumption.

	We  assume the stable-unit treatment value assumption (SUTVA), i.e., there is only one version of the treatment and the potential outcome $Y(z)$ of an individual does not depend on another individual's received treatment, as it is the case when participants' outcomes interfere with one another. \cite{rosenbaum1983central} 
	To identify causal estimands of interest, we assume that $Y(0)$ and $Y(1)$ are conditionally independent of $Z$ given the vector of covariates ${X}$, i.e., $E[Y(z)|X]=E[Y(z)|X,Z=z],$ $z=0,1$ (unconfoundness assumption) for ATE. For ATT (resp. ATC), we only require that $E[Y(0)|X]=E[Y(0)|X,Z=1]$ (resp. $E[Y(1)|X]=E[Y(1)|X,Z=0]$) and we do not need to impose any condition on $Y(1)$ (resp. $Y(0)$).
	
	We define the propensity score $e({x})=P(Z=1|{X=x})$, i.e., conditional probability of treatment assignment  given the observed covariates. Under unconfoundness assumption, Rosenbaum and Rubin demonstrate that the propensity score is a balancing score since $X\bigCI Z|e({X})$. This implies that,  for participants with the same propensity score, the distributions of their corresponding observed baseline covariates ${X}$ are similar regardless of their treatment assignment.
	\cite{rosenbaum1983central,rosenbaum1984reducing,rubin1997estimating} Therefore, instead of controlling for the whole vector of multiple covariates $X$ to estimate treatment effects, one can leverage this property of the propensity score $e({X})$ to derive unbiased estimators of the treatment effect.

	More often, the goal is to estimate the average treatment effect (ATE) defined as $\tau=\displaystyle E[\tau(X)]$  where  $\tau(x)$ is the conditional average treatment effect  $E[Y(1)-Y(0)|X=x]$. 
	In this paper, we investigate the weighted average treatment effect (WATE) 
	\allowdisplaybreaks\begin{align}\label{eq:estimand_gen}
		\tau_{g}&=\displaystyle \frac{\displaystyle E[g(X)\tau(X)]}{\displaystyle E(g(X))}
		=C^{-1}\!\displaystyle {\displaystyle\int \tau(x)f(x)g(x)dx},~~\text{with}~~ C={\displaystyle\int g(x)f(x)dx}
	\end{align} 
	where $f(x)$ represent the marginal density of the covariates with respect to a base measure 	 $\mu,$ which  we have equated to the Lebesgue measure, without loss of generality.
	The WATE estimand $\tau_g$ generalizes a large class of causal estimands.\cite{crump2006moving,crump2009dealing,hirano2003efficient,li2018balancing}  The selection function $g$ specifies the target subpopulation as well as the treatment effect estimand  of interest and helps characterize the related weights. \\
	In the literature,  ATE, ATT and ATC are often defined  as 
	\allowdisplaybreaks\begin{align*}
		\tau_{\text{ATE}}& = E[Y(1)-Y(0)] = E[Y(1)]-E[Y(0)] \\
		\tau_{\text{ATT}}& = E[Y(1)-Y(0)|Z=1] = E[Y|Z=1]-E[Y(0)|Z=1] \\
		\tau_{\text{ATC}}& =  E[Y(1)-Y(0)|Z=0]  = E[Y(1)|Z=0]-E[Y|Z=0].
		\end{align*} 
	 Using simple algebra, we can show that they correspond to $\tau_{g}$ for $g(X)=a+be(X)$, where $(a,b) =(1,0)$ for ATE,   $(0,1)$ for ATT and $(1, -1)$ for ATC, respectively. \cite{hirano2003efficient}
	
	The expectation in \eqref{eq:estimand_gen} is taken over the population of interest and $\tau_{g}$ simplifies to $\tau$ when $g\equiv 1$.  The product $f(x)g(x)$ represents the target population density, where the function $g(X)$, which we refer to as the selection function, is a known function of the covariates \cite{hirano2003efficient} and can be modeled as $g(X; \beta)$ with  parameters  $\beta$.	
	\subsection{Point estimation}\label{sec:pt_estimation}
	The propensity score $e(X)$ is usually unknown; we estimate it by postulating a model
	$e({X};{\beta})=P(Z=1|{X}; {\beta})$, for some parameter vector $\beta$.  	In this paper, we will estimate $e(X)$ using a logistic regression model parametrized by the coefficients $\beta = (\beta_0,\ldots, \beta_p)'$ such that $e(X;{\beta})=(1+\exp(-X\beta))^{-1}$. The parameter $\beta$ is estimated via the maximum likelihood estimation using the score function  $\psi_{{\beta}}({X}_i, Z_i)=[Z_i-e({X_{i}};{\beta})]X_{i}$. Other methods to estimate $e(X)$  including generalized additive models \cite{woo2008estimation} and machine learning techniques such as generalized boosted regression models, random forest, classification and Bayesian additive regression trees. \cite{lee2010improving, kern2016assessing, stuart2017generalizing}
	
	Let $\widehat \beta$ a consistent estimator $\beta$ and consider $\widehat e(x) =e(x;\widehat \beta)$ an estimator  of $e(X;{\beta})$. 	
	We define the weights	$\widehat w_z(x_i) =\widehat t(z, Z_i, x_i)\Big/\displaystyle\sum_{i=1}^{N}\widehat t(z, Z_i, x_i)$,
 	where $\widehat t(0, Z, x) =(1-Z)\widehat g(x)/(1-\widehat e(x))$,  $\widehat t(1, Z, x) ={Z}\widehat g(x)/{\widehat e(x)},$  and   $\widehat  g(x)= a + b\widehat e(x)$,  for  $z, a, b \in \{0, 1\}$. 	
	The weights $\widehat w_z(x)$ are expected to be finite for every participants---such a condition is often called the positivity assumption.  For ATE, it imposes that $0<\widehat e(x)<1$. For ATT, $\widehat w_0(x)$ is defined if $\widehat e(x)<1$, while for ATC   $\widehat w_1(x)$ is defined if $\widehat e(x)>0.$
	
	An estimator of the treatment effect $\tau_{g}$ is given by	
	\allowdisplaybreaks\begin{equation}\label{eq:wgted_est}
		\widehat\tau_{g}= \displaystyle\sum_{i=1}^{N}\left(\widehat  w_1(x_i)-\widehat w_{0}(x_i)\right) Y_i.
	\end{equation} 
	
	We can also determine an estimator of $\tau_{g}$ for ATT and ATC by postulating parametric outcome regression (OR) models $m_z(X; \alpha_z)$ for the conditional mean models $m_z(X)=E(Y|Z=z, X)$, $z=0,1$. This requires to impute the potential outcome $Y(0)$ (resp. $Y(1)$)  for treated (resp. control) participants through corresponding fitted values, i.e., 
	\allowdisplaybreaks\begin{align}\label{eq:est_reg}
		\widehat\tau_{_\text{ATT}} &=\displaystyle\sum_{i=1}^{N}\widehat w_{1}(x_i){[Y_i-\widehat m_{0}(x)]}~~~\text{and}~~~
		\widehat\tau_{_\text{ATC}} =\displaystyle\sum_{i=1}^{N}\widehat w_{0}(x_i)[\widehat m_{1}(x)-Y_i]
	\end{align}
	where $\widehat m_z(X)= m_z(X; \widehat\alpha_z)$ and $\widehat\alpha_z$ is an estimator of the parameter $\alpha_z= (\alpha_0, \dots, \alpha_q).$	
			
	Note that in both \eqref{eq:wgted_est} and \eqref{eq:est_reg}, we focus on normalized weights within treatment groups, i.e.,  $\displaystyle\sum_{i=1}^{N} \widehat w_z(x_i)=1$, $z=0, 1.$ When $g(x) =1,$ the ATE estimator $\widehat \tau_{g}$ in \eqref{eq:wgted_est} correspond to the modified Horvitz-Thompson IPW estimator, also known as the Hajek's estimator. \cite{hajek1971comment} Other estimators of ATT and ATC, without normalized weights in \eqref{eq:wgted_est}, have  been used.\cite{moodie2018doubly,mercatanti2014debit}  The choice of normalized weights in this paper stem from the fact they often lead to improved finite sample  properties.\cite{busso2009finite, busso2014new} 
	
	The performance of the  weights $w_z(x)$ can be assessed via the effective sample size (ESS), estimated by 
	\allowdisplaybreaks\begin{align*} 
		\widehat{ESS}= \left(\displaystyle\sum_{i=1}^{N} \widehat w(x_i)^2\right) ^{-1} \left( \displaystyle\sum_{i=1}^{N} {\widehat w(x_i)}\right)^{2},~~ \text{where} ~~\widehat w(x_i)=z_i\widehat w_1(x_i)+(1-z_i)\widehat w_0(x_i).
	\end{align*}  
	The ESS  provides the approximate number of independent observations drawn from a simple random sample needed to obtain similar amount of information (in terms of sampling variation) in estimating $\tau_g$ than that of the weighting observations. 	Thus, by estimating the design effect $DE= N^{-1} ESS$, we can have an indication of the potential loss of precision. 
	
	For ATT (resp. ATC) we only (re)weight the control (resp. treated) participants while assigning a weight of 1 to the other group. Therefore, the assessment via ESS is often limited to that weighted group.\cite{mccaffrey2004propensity}  Thus, for ATT (resp. ATC), we only use $\widehat w(x_i)=\widehat w_0(x_i)$ (resp. $\widehat w(x_i)=\widehat w_1(x_i)$) and $DE= N_1^{-1} ESS$ (resp. $DE= N_0^{-1} ESS$), where $N_z=\displaystyle\sum_{i=1}^{N} Z_i^z(1-Z_i)^{1-z}, ~z \in \{0,1\}$.	
	 The ESS is an important measure, especially when there is practical violation of the positivity assumption, i.e., when some $\widehat e(x)\approx 1$ for ATT or   when  we estimate the ATC and  some $\widehat e(x)\approx 0$.   
	 	 
	Alternatively, one can also estimate directly the variance inflation, which characterizes  the precision loss due to weighting\cite{kish1985survey}
	\allowdisplaybreaks\begin{align*}
		\widehat{\text{VI}}=\frac{N_1N_0}{N}  \sum_{z = 0}^1 \left[ \left( \displaystyle\sum_{i=1}^{N_z} {\widehat w_z(x)}\right)^{-2}\!\!\displaystyle\sum_{i=1}^{N_z} \widehat w_z(x_i)^2\right].
	\end{align*} 
	The smaller the $\widehat{\text{VI}},$ the better; higher values would indicate that the weights provide a pseudo-population that is far from what we can expect had we drawn participants randomly to be assigned to one or the other treatment group---which can lead to biased point estimates and high related variance.  
	
	Guided by subjet-matter knowledge, the research questions, and the data at hand, the choice of covariates $X$ to include in the models $e({X};{\beta})$ or  $m_z(X; \alpha_z)$, $z=0, 1$ should privilege good fit and prediction: they must only be true confounders  (i.e., common causes of the treatment--outcome relationship) and prognostic variables  (i.e., those that are related to the outcome $Y$)---regardless of whether the latter are significantly related to the treatment $Z$ or not.\cite{kang2007demystifying,vanderweele2019principles,tsiatis2007semiparametric}  
	
	Beyond the SUTVA and unconfoundness assumptions of Section \ref{sec:estimation} as well as the positivity assumption of Section \ref{sec:pt_estimation}, consistency of the estimators \eqref{eq:wgted_est} and \eqref{eq:est_reg} depends on a different set of conditions. The estimators \eqref{eq:wgted_est} are consistent only when the propensity scores $\widehat e_z(x)$ are correctly estimated, while the regression-based estimators \eqref{eq:est_reg} rely on the regression models $\widehat m_z(x)$ being correctly specified. Otherwise, these estimators would be biased. In practice, since we can never be really sure which of these models is correctly specified, it may be hard to pinpoint in advance the most desirable approach that leads to a reliable estimate  of $\tau_{g}$.

	
	\subsection{Doubly robust estimators}\label{sec:dr_estimation}
	 
	 Instead of choosing between a PS or an OR  parametric model to specify, the IPW augmentation framework allows us to judiciously leverage the two models and obtain a doubly robust estimator of $\tau_{g}$. Such an estimator remains consistent if at least one of the two models is correctly specified. \cite{tsiatis2007semiparametric,moodie2018doubly} 	 
	 Concurrent with the normalized estimators of Section \ref{sec:pt_estimation}, the idea is to derive a biased-corrected estimator of the conditional expectation $E[Y(z)|Z=1-z]$, $z=0, 1$  through a clever combination of weighted averages of OR and the residuals $Y-m_z(X; \alpha_z)$. Indeed, 
	$$E[Y(z)|Z=1-z]=E[m_{z}(X)|Z=1-z] + E[Y(z)-m_{z}(X)|Z=1-z], ~z=0, 1,$$
 it can be estimated by 
	$ \displaystyle\sum_{i=1}^{N}\widehat w_{1-z}(x_i)\widehat m_z(x_i)+\displaystyle\sum_{i=1}^{N}\widehat w_z(x_i)[Y_i-\widehat m_{z}(x_i)].
	$
	
	Hence, the doubly robust estimator for the ATT and ATC  are, respectively
	\allowdisplaybreaks\begin{align} \label{eq:dr_est}
	\widehat  \tau_{_{\text{ATT},\,\text{DR}}}	& =\displaystyle\sum_{i=1}^{N}\left(\widehat  w_1(x_i)-\widehat w_{0}(x_i)\right) [Y_i-\widehat m_0(x_i)];\\
	\widehat  \tau_{_{\text{ATC},\,\text{DR}}}	& =\displaystyle\sum_{i=1}^{N}\left(\widehat  w_1(x_i)-\widehat w_{0}(x_i)\right) [Y_i-\widehat m_1(x_i)].\nonumber 
	\end{align}
	These two estimators share the strength of the estimators \eqref{eq:wgted_est} and \eqref{eq:est_reg}, while mitigating some of their limitations.   When the propensity score and the regression models are correctly specified, the doubly robust estimator can  also attain the  efficiency bound. \cite{hirano2003efficient,tsiatis2007semiparametric} Nevertheless, there is a bias--variance tradeoff that must be considered when one of the models happens to be correctly specified: the weighting estimators \eqref{eq:wgted_est} with a correctly specified PS model or the  regression-based estimators \eqref{eq:est_reg} with a correctly specified regression models will be more efficient than the  doubly robust estimators \eqref{eq:dr_est}. \cite{kang2007demystifying}
	
	It is worth mentioning that double robustness pertains only to the specifications of the models $e({X};{\beta})$ and $m_z(X; \alpha_z)$. The related consistency of a doubly robust estimator may not hold in the presence of extreme weights, which often occurs when the positivity assumption $e(X)<1$ for ATT  (resp. $e(X)>0$ for ATC) is violated.\cite{kang2007demystifying,kang2016practice} 
	
	Combining the propensity score and regression models is also the building block of the targeted maximum likelihood estimation (TMLE).
	\section{Variance estimation}\label{sec:variance_estimation}
	
	The doubly robust estimators $\tau_{_{\text{ATT},\,\text{DR}}}$ and $\tau_{_{\text{ATC},\,\text{DR}}}$ of the previous section are calculated in  two steps. First, we calculate the so-called nuisance parameters $\widehat \beta$ and $\widehat \alpha_z$ and estimate $\widehat e({X})$ and $\widehat m_z({X})$ from the parametric  propensity score and the regression models $e({X};{\beta})$ and $m_z(X; \alpha_z)$. Then, we plug these estimates into \eqref{eq:dr_est} to determine either $\widehat  \tau_{_{\text{ATT},\,\text{DR}}}$ or $\widehat  \tau_{_{\text{ATT},\,\text{DR}}}$ as needed. It is just natural that how we estimate the nuisance parameters $\widehat \beta$ and $\widehat \alpha_z$ also affects  our   estimates of  $\widehat  \tau_{_{\text{ATT},\,\text{DR}}}$ and $\widehat  \tau_{_{\text{ATT},\,\text{DR}}}$. Therefore, we must factor in the uncertainty related with estimating of nuisance parameters $\widehat \beta$ and $\widehat \alpha_z$ into how we  eventually estimate the variance of the estimators $\widehat  \tau_{_{\text{ATT},\,\text{DR}}}$ and $\widehat  \tau_{_{\text{ATT},\,\text{DR}}}$.
	\subsection{Sandwich variance estimation}
	In this section we present variance estimation of the ATT and ATC using M-theory (see Stefanski and Boos \cite{stefanski2002calculus}). The variance of an  estimator $\widehat\tau_{g}=c'\theta$, for some vector $c$, is derived by solving an estimating equation $\displaystyle 0=\displaystyle\sum_{i=1}^{N} \Psi_\theta({X}_i, Z_i, Y_i)$, with respect to $\theta$, for some matrix $ \Psi_\theta({X}_i, Z_i, Y_i).$ 
	
	Starting with $E[\Psi_\theta({X}_i, Z_i, Y_i)]=0$, it can shown that $\widehat\theta\overset{p}{\longrightarrow}  \theta $ when  $N\longrightarrow \infty$, under some  regularity conditions. 
	\cite{stefanski2002calculus,bodory2020finite} Thus, by Slutsky's theorem, $ \widehat\theta$ is consistent for $\theta$. In addition,  $\sqrt{N}(\widehat{\theta}-{\theta})\overset{d}{\longrightarrow}N({0}, \Sigma({\theta})),$ with $ \Sigma({\theta})=A(\theta)^{-1}B(\theta)\{A(\theta)'\}^{-1}$. A consistent estimator of variance-covariance matrix $\Sigma({\theta})$ of $\widehat{\theta}$ is  $ \widehat \Sigma({\widehat\theta})=A_N(\widehat\theta)^{-1}B_N(\widehat\theta)\{A_N(\widehat\theta)'\}^{-1}$, where  $A(\theta), B(\theta),$ $ A_N(\widehat\theta)$ and $B_N(\widehat\theta)$ are the following matrices: 
	\begin{align*}
		A_N(\widehat{\theta})&=\frac{- 1}{N}\displaystyle\sum_{i=1}^{N}\frac{\partial \Psi_\theta({X}_i, Z_i, Y_i)}{\partial{\theta'}}\Big|_{\theta={\widehat\theta}}; ~~B_N(\widehat{\theta})= \frac{1}{N}\displaystyle\sum_{i=1}^{N}\Psi_\theta({X}_i, Z_i, Y_i)\Psi_\theta({X}_i, Z_i, Y_i)'\big|_{\theta=\widehat\theta};
		\\
		A(\theta)&=\lim_{N\rightarrow\infty}A_N(\widehat\theta)  ~\text{and}~~
		B(\theta)=\lim_{N\rightarrow\infty}B_N(\widehat\theta)=E\left[ \Psi_\theta({X}_i, Z_i, Y_i)\Psi_\theta({X}_i, Z_i, Y_i)'\right]. 
	\end{align*} 
 Therefore, an estimator of the variance of $\widehat\tau_{g}=c'\theta$ is given by $ c'\widehat \Sigma({\widehat\theta})c.$
 
	We provide in the Supplemental Material \ref{sec:sandvariance_estimation} examples of the matrices $ \Psi_\theta({X}_i, Z_i, Y_i),$  $A_N(\widehat{\theta})$ and $B_N(\widehat{\theta})$ for the non parametric estimators \eqref{eq:wgted_est} as well as for the doubly robust estimators \eqref{eq:dr_est}, when the propensity score and regression models are estimated using the logistic and linear regression models, respectively.
	
	We should emphasize that the above variance estimation works only if the matrices  $\Psi_\theta({X}_i, Z_i, Y_i)$ and $\Psi_\theta({X}_i, Z_i, Y_i)\big|_{\theta=\widehat\theta}$ are of full rank, i.e., $B_N({\theta})$ and $B_N(\widehat{\theta})$ are non-singular. This is  the case when the propensity and the outcome models  are estimated via logistic and linear regression models with appropriately selected covariates, respectively. However,  the variance estimation may not hold when we postulate other types of models or use some machine learning techniques that does not guarantee the non-singular conditions for $B_N({\theta})$ and $B_N(\widehat{\theta})$.
	\subsection{Wild bootstrap variance estimation}
	While studying asymptotic properties of estimators is extremely important, what matters most to consider in practice  is the behaviours of these estimators under finite sample size. For variance estimation, there have been some interest in the use of bootstrap variance estimation, especially when the sample size is small or moderate and there are intermediate steps to estimate nuisance parameters---as was also remarked in the context of propensity score matching.\cite{abadie2008failure} 
		
	Unfortunately, what makes the strength of standard bootstrap algorithm can also be a drawback when we estimate the variance of ATT or ATC. \cite{abadie2008failure, hastie2009elements} For an efficient estimation of the variance, the standard bootstrap requires that we draw samples directly from the original data with replacement. However, this can  lead to a small fraction of observations from either treatment group in some bootstrap iterations.  Running  propensity score model on such lopsided bootstrap iterations can exacerbates the lack of positivity, crucial to identifying the point estimate. On the other hand, since the models $m_z(x)$ are run on the subset $Z=z$, $z=0,1$, whenever such a subset is small, estimation of $\widehat m_z(x)$ hinges on extrapolation via the postulated parametric models, which may no longer be reliable.\cite{zhou2020propensity} Therefore, the impact of the nuisance parameters from these bootstrap iterations can drastically affect bootstrap estimates of $\widehat \tau_g$ and thus the overall bootstrap variance estimation.
	
	A wild bootstrap procedure approximates the sampling distributions of the estimators without disrupting the proportions of treated and control participants in each bootstrap sample. Hence, as alternative to the standard bootstrap procedure, we propose the use of wild bootstrap  to estimate the variance of ATT and ATC. \cite{davidson2008wild,shao2010dependent,chernozhukov2018sorted} Therefore, instead of sampling  with replacement from the original data, we simply perturb the efficient influence functions (which we define below) of these estimators and estimate their variances accordingly.
	
	\subsubsection{Influence functions}
	
		The estimators $\widehat\tau_{g}$ of ATT and ATC are regular and asymptotically linear (RAL) estimators, i.e.,  $\sqrt{N}(\widehat \tau_g-\tau_g)=\displaystyle \frac{1}{\sqrt{N}}\displaystyle \sum_{i=1}^{N}\phi_g(Z_i, X_i, Y_i) + o_p(1)$, under some regularity conditions,
	where $\phi_g(Z, X, Y)$ is the influence function of $\tau_g$, for which $E[\phi_g(Z, X, Y)]=0$  and $Var[\phi_g(Z, X, Y)]=\displaystyle \Sigma_g = E[\phi_g(Z, X, Y) \phi_g(Z, X, Y) ']$. Thus, 	$\sqrt{N}(\widehat \tau_g-\widehat\tau_g)\overset{d}{\longrightarrow} N(0, \Sigma_g)$ as  $N\longrightarrow\infty$, assuming some regularity conditions. \cite{tsiatis2007semiparametric,hirano2003efficient, bodory2020finite} 	

	The efficient influence functions of ATT and ATC are given, respectively, by \cite{hahn1998role}	
	\begin{align*}
		\phi_{_\text{ATT}} &=\displaystyle\frac{1}{p} \left[\frac{(Z-e(X))}{1-e(X)}(Y-m_0(X))-Z\tau_{_\text{ATT}}\right], ~\text{where}~p=P(Z=1);\\
		\phi_{_\text{ATC}} &=\displaystyle\frac{1}{1-p}  \left[\frac{(Z-e(X))}{e(X)}(Y-m_1(X)) -(1-Z)\tau_{_\text{ATC}}\right].
	\end{align*}
	Moreover,
	$\sqrt{N}(\widehat \tau_g-\tau_g)\overset{d}{\longrightarrow} N(0, \Sigma_g) $  as $ N\longrightarrow\infty.$ We can use the RAL property to reproduce the sampling distribution of $\sqrt{N}(\widehat \tau_g-\tau_g)$ via some perturbations. 
	 \subsubsection{Wild bootstrap algorithm}
	Consider $\widehat \phi_g(Z, X, Y)$ an estimator of  $\phi_g(Z, X, Y)$, where we replace $\tau_g$ by its related estimator $\widehat\tau_g$, given in Section \ref{sec:dr_estimation}.   
	We generate random vectors $\xi = (\xi_1, \dots, \xi_\text{N})$ of independent identically distributed random variables---with mean 0 (or 1), variance 1, and $E(\xi_i^3)<\infty $, where $\xi$ does not depend on the observed data $O=\{ (Z_i, X_i, Y_i): i=1\dots, N \}$. Define $\widehat \tau_g^* = \widehat \tau_g+ \displaystyle {N}^{-1}\sum_{i=1}^{N}\xi\widehat \phi_g(Z_i, X_i, Y_i)$. We have, 
	$\widehat \Delta_g  =\sqrt{N}(\widehat \tau_g^*-\widehat\tau_g)=\displaystyle {N}^{-1/2}\displaystyle \sum_{i=1}^{N}\xi\widehat \phi_g(Z_i, X_i, Y_i) + o_p(1)\Rightarrow \widehat \Delta_g \overset{d}{\rightarrow} N(0, \Sigma_g^*)$ as
	$ N\longrightarrow\infty$,  under some regularity conditions,  where $\Sigma_g^* = E\left[\xi\widehat\phi_g(Z_i, X_i, Y_i)\widehat\phi_g(Z_i, X_i, Y_i)' \xi'\right].$		
	Moreover, the empirical distribution of $\widehat \Delta_g$, conditional on the observed data, approximates the sampling distribution of $\sqrt{N}(\widehat \tau_g-\tau_g).$ This result establishes the validity of the bootstrap procedure (see Chernozhukov et al.,\cite{chernozhukov2018sorted} for details) whose algorithm  we now describe. Let $r=1,\dots, R$, where  $R=$ number of bootstrap replicates.
	\begin{enumerate}\label{eq:algorithm}
		\item Generate the random vector $\xi = (\xi_1, \dots, \xi_\text{N})$, under Rademacher distribution with $P(\xi_i=-1) = 0.5$ and $P(\xi_i=1) = 0.5$ or under the standard exponential distribution with mean 1 and variance 1;
		\item Compute $\widehat \tau_g^*(r) = \widehat \tau_g+ \displaystyle {N}^{-1}\sum_{i=1}^{N}\xi\widehat \phi_g(Z_i, X_i, Y_i)$ and  $\widehat \Delta_g(r)  = \sqrt{N}(\widehat \tau_g^*(r)-\widehat \tau_g)$;
		\item Repeat the steps 1--2, $R$ times;
		\item Calculate $ \widehat \Sigma_g^{1/2}$  an estimate of $ \Sigma_g^{1/2}$ as $ \widehat \Sigma_g^{1/2}= \displaystyle {\text{IQR}(\widehat \Delta_g)}/{(z_{0.75}-z_{0.25})}$, i.e., the interquartile range of $\widehat \Delta_g(r)$ across all the $R$ bootstrap replicates,  rescaled with the interquartile range $z_{0.75}-z_{0.25} = 1.3489795$ of the standard normal distribution.
		\item Calculate the $100(1-\alpha)\%$  confidence interval of  $\tau_g$ as  $\widehat \tau_g \pm {N}^{-1/2}{z_{(1-\alpha/2)}}\widehat \Sigma_g^{1/2}$. 
	\end{enumerate}
	 {\it Remarks}:	
	 \begin{itemize}
	 \item We have used, in step 4, the bootstrap procedure based on the t-statistic and derived, in steps 5 and  6, the confidence interval using quantiles. We can also estimate the confidence intervals and even the p-values using either the asymptotic distribution of the t-statistic or the percentiles of $|\widehat \tau_g|\widehat \Sigma_g^{-1/2}$. 
	 \item Since,  $\Sigma_g^* = E\left[\xi\widehat\phi_g(Z_i, X_i, Y_i)\widehat\phi_g(Z_i, X_i, Y_i)' \xi'\right],$	we  can also directly approximate the variance $\Sigma_g^*$ by   $\widehat \Sigma_g^*=R^{-1}\displaystyle \sum_{r=1}^{R}\widehat \Delta_g(r)  \widehat \Delta_g (r)'.$
	 
	 \item Bootstrap estimators offer the opportunity to reduce bias by using  the bias-corrected estimator  $2\widehat \tau_g - \displaystyle {\widehat\tau_g^*}/{R}$ of $\tau_g$, with the confidence interval given by  $2\widehat \tau_g - \overline{\widehat\tau_g^*}\pm \displaystyle {z_{(1-\alpha/2)}}{N}^{-1/2}\widehat \Sigma_g^{1/2},$ where $\overline{\widehat\tau_g^*}=\displaystyle \displaystyle \sum_{r=1}^{R} \widehat\tau_g^*(r)/{R}$.
	 \end{itemize}

	\section{Simulation}\label{sec:simulations}
	We use Monte-Carlo simulations to evaluate the different aforementioned methods, following the data generating process of Li and Li. \cite{li2021propensity} We first simulated a superpopulation of $10^6$ individuals with desired proportions of treated participants $P(Z=1)$ to calculate the true estimands under heterogeneous treatment effect. Then, we generated $M=1000$ samples of size $N=1000$ to assess the finite-sample performance of the variance estimation methods presented in Section \ref{sec:variance_estimation}, under different scenarios, specified below. \textcolor{black}{For comparison, we also included variance estimation based on the standard bootstrap and the targeted maximum likelihood estimation (TMLE).}
	
	{\color{black} 
	We report the simulation results in terms of point estimation (Est), absolute relative percent bias (Bias), root mean squared error (RMSE), median standard error estimation (SE), empirical standard deviation (ESD), relative efficiency (RE), i.e., the ratio of the empirical variance to the corresponding variance estimation, and  coverage probability (CP) under heterogeneous treatment effects. 
	 	
	All bootstrap methods were based on $R = 1000$ replicates, i.e., for each generated sample $m=1,\dots, M$, we re-sampled with repetition (standard bootstrap) from the original sample ($m$) or perturbed (wild bootstrap) the influence function to obtain  $R$ bootstrap replicates. To estimate the variance under standard bootstrap, we first re-estimated the propensity scores and weights in each bootstrap sample, then calculated the causal effects corresponding to each estimand. Finally, we derived the standard error as the standard deviation of the causal effects across $1000$ bootstrap replicates. For wild bootstrap, we followed the algorithm given in Section \ref{eq:algorithm} to estimate the variance.}
	 
	\textcolor{black}{For TMLE, we used the R package ``tmle'' and specified the propensity score and outcome models as generalized linear models, instead of defaulting to SuperLerner with more flexible estimation.\cite{gruber2012tmle} Therefore, to ensure comparability across different methods, we did not use the data-adaptive features of TMLE to optimize the choice of covariates to include in the models nor its potential to truncated extreme propensity score weights at a specific threshold.}
	
	\subsection{Data generating process (DGP)}\label{subsec:DGP}
	
	The DGP consists of 7 random variables $X_4\sim \text{Bern}(0.5), X_3\sim \text{Bern}(0.4+0.2X_4)$, $(X_1, X_2)'\sim N(\boldsymbol{\mu, \Sigma})$, $X_5 = X_1^2,$ $ X_6 = X_1X_2,$  $ X_7 = X_2^2$, where $
	\boldsymbol \mu = (X_4-X_3  + 0.5X_3X_4, X_3-X_4 + X_3X_4)'
	$ and $
	\boldsymbol \Sigma = 
	X_3\begin{pmatrix}
		1 & 0.5 \\
		0.5 & 1
	\end{pmatrix} + $  
	$(1-X_3)\begin{pmatrix}
		2 & 0.25 \\
		0.25 & 2
	\end{pmatrix}
	$. 
	{\color{black} We generated the treatment indicator  $Z\sim \text{Bern}\left(  (1+\exp(-X\beta))^{-1}\right) $ from a logistic regression model, where  $X = (\mathbbm{1}, X_1, \dots, X_7)$ and $\mathbbm{1}=(1,\dots, 1)'$. For the first three models, we used $\beta=(\beta_0, 0.3, 0.4, 0.4, 0.4, -0.1, -0.1, 0.1)'$,  with $\beta_0 = -2.17, -0.78,$ and $0.98$ respectively. The fourth model was generated with $\beta= (0.2, 1.0, -0.9, -0.9, 0.9, 0.15, 0.15, -0.2)'$. Through these four models, we simulated different scenarios that arise in practice, including different proportions of treated participants (Models 1--3) and poor overlap along with extreme IPW weights (Model 4). We further investigated Model 1 under two small sample sizes ($N = 100$ and $N=50$), which we labeled as Model 5a and Model 5b (for the sake of exposition), which represent scenarios where the close-form sandwich variance estimation may not work. The full simulation details and different model characteristics can be found in the Supplemental Material \ref{appendix2}.  Figure \ref{fig:PS_dist}, for instance,  shows the distribution of estimated propensity scores under our second model, Model 2, which corresponds to $\beta = (-0.78, 0.3, 0.4, 0.4, 0.4, -0.1, -0.1, 0.1)'$.} 

	\begin{figure}
		\centering
		\includegraphics[width=0.6\textwidth]{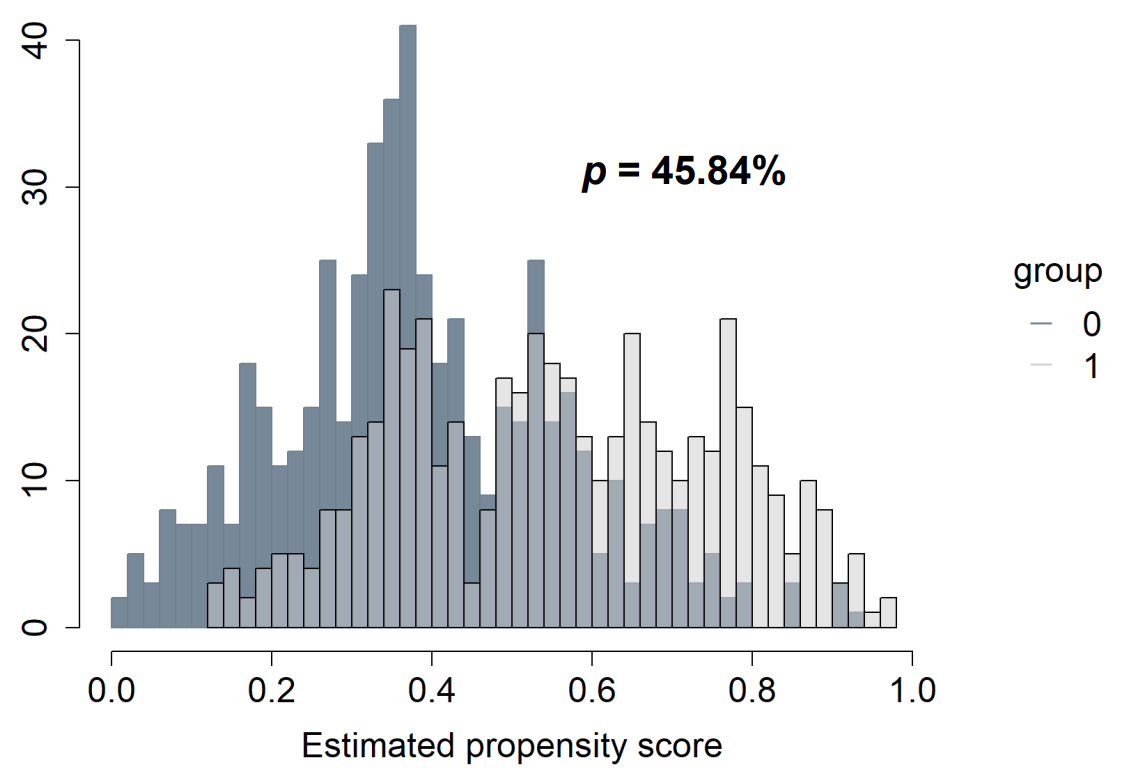}
		\caption{Estimated propensity scores for Model 2}\label{fig:PS_dist}
	\end{figure}

	Finally, we generated the observed outcome $Y=ZY(1)+(1-Z)Y(0)$, where $Y(0) = 0.5+X_1+0.6X_2+2.2X_3-1.2X_4+(X_1+X_2)^2 + \varepsilon$ and  $Y(1) = Y(0) + \delta(X)$, for $\varepsilon\sim N(0, 4)$. 	
	We considered both the constant treatment effect $\delta(X)=4$ and the heterogeneous treatment effect  $\delta(X)=4+3(X_1+X_2)^2+X_1 X_3.$ The true values of ATT and ATC for the heterogeneous treatment effects are shown in Table \ref{tab:truth} in the Supplemental Material \ref{appendix2_PS}. 
	

To evaluate the finite-sample performance of the proposed variance estimators, we also considered the cases where the PS model or the outcome regression (OR) models are misspecified.
In misspecified models some covariates were omitted; we excluded the quadratic and interaction terms $X_5$,  $X_6,$ and $X_7$  in the misspecified PS model  and  we excluded $(X_1+X_2)^2$ in the misspecified OR models, including in $\delta(X)$. 

	\subsection{Results}\label{subsec:sim-results}
\subsubsection{Main results} 
In this section, we present in Table \ref{sim-result-main}  the results under heterogeneous treatment effect from Model 2, where the proportion of the treated participants is $45.84\%$, overlap of the propensity score in two groups is good, and no extreme weights exist.

\begin{table}\footnotesize 
	\begin{threeparttable}
		\centering
		\caption{Simulation results from Model 2 (with $p=45.84\%$), under heterogeneous treatment effect}
		\label{sim-result-main}
		\begin{tabular}{llcccccccccccc}
			\toprule	
			& &\multicolumn{6}{c}{PS and OR models correctly specified} & \multicolumn{6}{c}{PS model correctly specified} \\\cmidrule(lr){3-8}\cmidrule(lr){9-14} 
			Est. & Method & Bias & RMSE & SE & ESD & RE & CP & Bias & RMSE & SE & ESD & RE & CP \\ 
			\midrule
			\multirow{5}{*}{ATT} 
			& Sand. & 0.02 &  0.93 & 1.26 & 0.93 & 0.55 & 0.99 & 0.29 & 1.04 & 1.33 & 1.04 & 0.60 & 0.99 \\ 			
			& WB R  & 0.02 & 0.93 & 0.92 & 0.93 & 1.03 & 0.95 & 0.29 & 1.04 & 1.32 & 1.04 & 0.62 &  0.99 \\  
			& WB E  & 0.02 & 0.93 & 0.91 & 0.93 & 1.05 & 0.94 & 0.29 & 1.04 &  1.30 & 1.04 & 0.63 & 0.98 \\ 
			& Std.Boot & 0.02 & 0.93 & 0.92 & 0.93 & 1.03 & 0.95 & 0.29 & 1.04 & 1.01 & 1.04 & 1.06 & 0.95 \\ 
			& TMLE & 0.00 & 0.93 & 0.92 & 0.93 & 1.03 & 0.95 & 1.24 & 1.11 & 1.23 & 1.09 & 0.79 & 0.97 \\ 

			\addlinespace
			\multirow{5}{*}{ATC}
			& Sand. & 0.00 & 0.81 & 1.05 & 0.81 & 0.59 & 0.98 & 3.16 & 3.83 &  2.81 & 3.80  & 1.83 &  0.93 \\  
			& WB R & 0.00 & 0.81 & 0.85 & 0.81 & 0.90 & 0.95 & 3.16  & 3.83 & 3.19 & 3.80  & 1.42 &  0.93 \\  
			& WB E & 0.00 & 0.81 & 0.84 & 0.81 & 0.92 & 0.95 & 3.16  & 3.83 & 2.80 & 3.80  & 1.84 & 0.92 \\
			& Std.Boot & 0.00 & 0.81 & 0.78 & 0.81 & 1.06 & 0.94 & 3.16 & 3.83 & 1.98 & 3.80 & 3.68 & 0.80 \\ 
			& TMLE & 0.33 & 0.80 & 0.77 & 0.79 & 1.06 & 0.94 & 1.68 & 2.56 & 1.93 & 2.54 & 1.73 & 0.90 \\ 
			\addlinespace 
			& &\multicolumn{6}{c}{PS and OR models misspecified} & \multicolumn{6}{c}{OR model correctly specified} \\\cmidrule(lr){3-8}\cmidrule(lr){9-14} 
			&  & Bias & RMSE & SE & ESD & RE & CP & Bias & RMSE & SE & ESD & RE & CP \\ 
			\midrule
			\multirow{5}{*}{ATT}  & Sand. & 3.83 & 1.47 & 1.54 & 1.30 & 0.70 & 0.96 & 0.02 & 0.93 & 1.26 & 0.93 & 0.55 & 0.99 \\ 
			& WB R  & 3.83 & 1.47 & 1.38 & 1.30 & 0.89 & 0.93 & 0.02 & 0.93 & 0.92 & 0.93 & 1.03 & 0.95 \\ 
			& WB E  & 3.83 & 1.47 &1.35 & 1.30 & 0.92 & 0.93 & 0.02 & 0.93 & 0.91 & 0.93  & 1.06 & 0.95 \\
			& Std.Boot & 3.83 & 1.47 & 1.26 & 1.30 & 1.05 & 0.90 & 0.02 & 0.93 & 0.92 & 0.93 & 1.02 & 0.95 \\ 
			& TMLE & 4.49 & 1.57 & 1.23 & 1.33 & 1.18 & 0.87 & 0.21 & 0.92 & 0.90 & 0.92 & 1.03 & 0.95 \\  
			\addlinespace 
			\multirow{5}{*}{ATC}& Sand. & 13.13 & 3.49 & 2.46 & 2.76 & 1.25 & 0.79 & 0.01 & 0.80 & 1.05 & 0.80 & 0.59 & 0.99 \\ 
			& WB R & 13.13 & 3.49 & 2.76 & 2.76 & 1.00  & 0.82 & 0.01 & 0.80 & 0.85 & 0.80  & 0.90 & 0.96 \\ 			  
			& WB E & 13.13 & 3.49 & 2.49 & 2.76 & 1.22 & 0.78 & 0.01 & 0.80 & 0.84 & 0.80  & 0.92 & 0.95 \\  	
			& Std.Boot & 13.13 & 3.49 & 2.11 & 2.76 & 1.70 & 0.72 & 0.01 & 0.80 & 0.78 & 0.80 & 1.04 & 0.94 \\ 
			& TMLE & 7.91 & 2.84 & 1.91 & 2.54 & 1.77 & 0.80 & 0.10 & 0.82 & 0.79 & 0.82 & 1.07 & 0.94 \\
			\bottomrule
		\end{tabular}
		\begin{tablenotes}
			\tiny
			\item  Est.: Estimand; ATT (resp. ATC): average treatment effect on the treated (resp. controls); Bias: absolute relative bias$\times 100$;  
			\item  RMSE: root mean squared error; SE: median of standard errors from proposed method; ESD: empirical standard deviation; RE: median of relative efficiencies; CP: coverage probability; Sand.: sandwich; WB R (resp. E) : wild bootstrap via Rademacher (resp. exponential) distribution; Std.Boot: standarded bootstrap; TMLE: targeted maximum likelihood estimation.
		\end{tablenotes}
	\end{threeparttable}
\end{table}	

	Overall, the simulation results in Table \ref{sim-result-main} indicate that, {\color{black} under a relatively large sample size ($N=1000$)}, {\color{black} the two wild bootstraps result in the best estimations in terms of standard error (SE), root mean squared error (RMSE), and coverage probability (CP)---mostly close to 0.95}. Only when PS and OR models are misspecified, some CPs are lower for the ATC (but usually between 0.75 and 0.95). The choice of different distributions for generating random vector in bootstrap does not affect much in variance estimation and constructing the confidence interval (CI). 
	
	 {\color{black}The standard bootstrap also performs well. Nevertheless, it is sometimes less robust to model misspecification compared to wild bootstrap. For example, whenever the outcome models are misspecified, the average CPs of the standard bootstrap variance estimation for ATC are often much lower than the nominal level of 0.95, while the wild bootstrap variance estimations perform better. The TMLE sometimes results in larger biases than the double-robust estimations of ATT and ATC, and is more sensitive to model misspecifications. Part of this poor performance is probably due to the non-use of the TMLE's data-driven iterative process of selecting the best models.\cite{van2013estimating}} Furthermore, for the finite sample size we considered, the sandwich variance estimation results in the largest coverage probability, and in many cases the CP is larger than 0.95 and close to 1. Thus, this method is the most conservative one in our simulations.  {\color{black} The results under constant treatment are similar to what we found in heterogeneous treatment effect aforementioned, where the two wild bootstrap methods have the overall best variance estimation for double-robust ATT and ATC with respect to the coverage probability, and more robust to model misspecifications.}
	 
	 \subsubsection{Additional models} 
	
	{\color{black} In addition to the results from Model 2, shown in Table \ref{sim-result-main}, we also provide a detailed presentation of the results and comments for Model 1--Model 5b in the Supplemental Material \ref{appendix2}. There, we deliberately simulated cases that help tease out characteristics of the proposed methods from different angles---small, medium, and large $p$ (Model 1, 2, and 3), presence of poor overlap and extreme IPW weights (Model 4), and sample size is small (Model 5a and 5b)---and ascertained the results under both constant and heterogeneous treatment effects. Plots depicting the overlap of the distributions of estimated propensity scores by treatment group, under these scenarios, are also presented in the Supplemental Material \ref{appendix2} (see Figures \ref{fig:ps_md123} and \ref{fig:ps_md4}.).
						
	 Congruent with literature, several patterns were observed when we varied the proportion of treated participants (Models 1-3). For instance, the point estimates were more biased when the OR models were misspecified than when the PS model was.\cite{kang2007demystifying} Having correctly specified OR models yielded results almost as good as having both PS and OR models correctly specified.  With a small $p$ (Model 1), the bias tend to be higher when estimating ATC than estimating ATT whereas for large $p$ (Model 3) it was completely the opposite---showing a dichotomy in estimating ATT with small $p$ versus  ATC estimation with a large $p$. From the results of Models 1 to 3, we found that overall the wild bootstrap methods were the  best with  regard to SE and CP, followed by the standard bootstrap in  many scenarios. The sandwich variance method was at times conservative (slightly larger SE and CP above 0.95), while the TMLE was biased in unexpected scenarios (all parametric models correctly specified or PS alone correctly specified). In other words, when we compare the different methods, we  reached a conclusion similar to that from of Model 2 under heterogenous treatment effect aforementioned: two wild bootstrap methods have the overall best variance estimation. 
 
	As we expected, the results from Model 4 (with $p=46.65\%$), which focuses on lack of overlap and extreme weights, yielded estimates with larger biases and higher SEs compared to the results from Model 2. Nevertheless, these results showed similar trends in favor the wild bootstrap and standard bootstrap methods. The impact of limited overlap and the presence of extreme weights was more pronounced in ATT estimation when at least one model was misspecified---under constant treatment effect. However,  when the treatment effect was heterogeneous slightly higher biases were  actually noticeable when we estimated  ATC. 
	Again bias and coverage probabilities were better when the OR models were correctly specified than when it was the PS, regardless whether the treatment effect was constant or not.  } 
	
	{\color{black} As proof of concept, we also considered scenarios with small sample sizes ($N=50$ and $N=100$) where the (large-sample) sandwich variance estimator might not work (model 5a and 5b). 
	Obviously, due to the small number of participants and low prevalence of treatment, sandwich variance estimation of ATC sometimes failed to return valid variance estimates
	(See Table \ref{tab:NAsmaller} in the Supplemental Material). As we explained in details in the Supplemental Material \ref{appendix2_PS}, the principal reason for such a high number of nonobtainable variance estimates 
	for ATC is the collusion of small sample size and the small proportion of treated participants, which led to singular estimated information matrices for a number of simulation data replicates.
		
	After excluding replicates where we could not estimate the variances, we summarized the results for $N=100$ (Model 5a) in Table \ref{sim-model5.effect}  in the Supplemental Material \ref{appendix2_VE}. In this case, we found that the standard bootstrap overall has the best CP and Bias, while the sandwich variance estimation and wild bootstrap, especially for ATC, did not always perform well. This result is expected because both sandwich variance estimation and wild bootstrap method rely on the large-sample property. 
	There were numbers of expected results we can point out to that are related to unstable models and lack of precision, even when some of the models (or all of them) were correctly specified. For instance, most coverage probabilities under ATC estimation were below 0.90, while they were often above 0.90 and sometimes close to 1 when we estimated ATT. Nevertheless, there were also some unexpected results that highlighted the impact of model specification.
	Under constant treatment effect, TMLE was biased when all the models were correctly specified and had a CP of 0.85 when estimating ATT. All the other methods had small biases and their CPs were equal to or above 0.90. For all the methods, there was a large bias when we estimated ATT (and even a larger bias when estimating ATC) whenever the outcome regression models were misspecified compared to when the propensity score was misspecified. All around, the standard bootstrap had  better standard errors and coverage probabilities compared to the other methods. TMLE was less biased throughout and had an acceptable CP while the other methods were more biased and did not always had a good CP. Similar behaviors were also noticed under heterogeneous treatment effect as well. Nevertheless, TMLE was more biased in estimating ATT or ATC when all the models were correctly specified and remained more biased when estimating ATT if at least one model was misspecified. 
			
	} 
	\section{Data illustration}\label{sec:examples}
	We consider data from the Study to Understand Prognoses and Preferences for Outcomes and Risks of Treatments (SUPPORT) on the use of right heart catheterization (RHC) diagnostic procedure during the initial care of hospitalized, critically ill patients\cite{connors1996effectiveness}. The objective of the study was to evaluate the effectiveness of RHC used during the first 24 hours stay in the intensive care unit (ICU) on patients' outcomes, including 30-day survival, cost of care,  and length of stay.   In total, 5735 hospitalized patients were enrolled in the study with 2184 (38\%) of them managed with RHC. 
	
	To illustrate the different variance estimation methods, we considered the effect of RHC on the patients' length of stay and used the 72 collected covariates  to estimate the propensity score and regression models, as previously done in the literature.\cite{mao2020flexible,hirano2001estimation}  The data are accessible on Vanderbilt Biostatistics website: \url{https://hbiostat.org/data/}.
	
		\begin{figure}[h]
		\centering
		\includegraphics[width=0.6\textwidth]{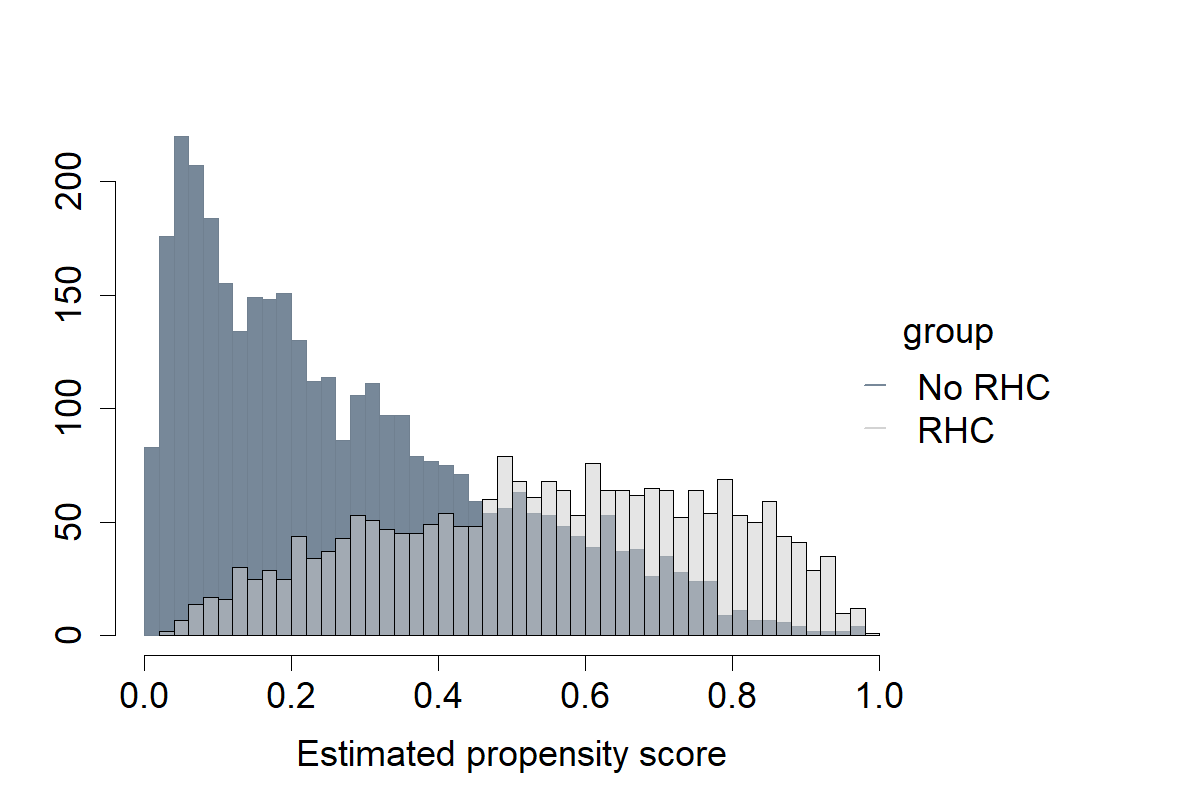}
		\caption{Propensity score estimates, based on the use of RHC in the SUPPORT Data}\label{Figure_PS.RHC}
	\end{figure}
	
	\begin{figure}
		\centering
		\includegraphics[width=0.6\textwidth]{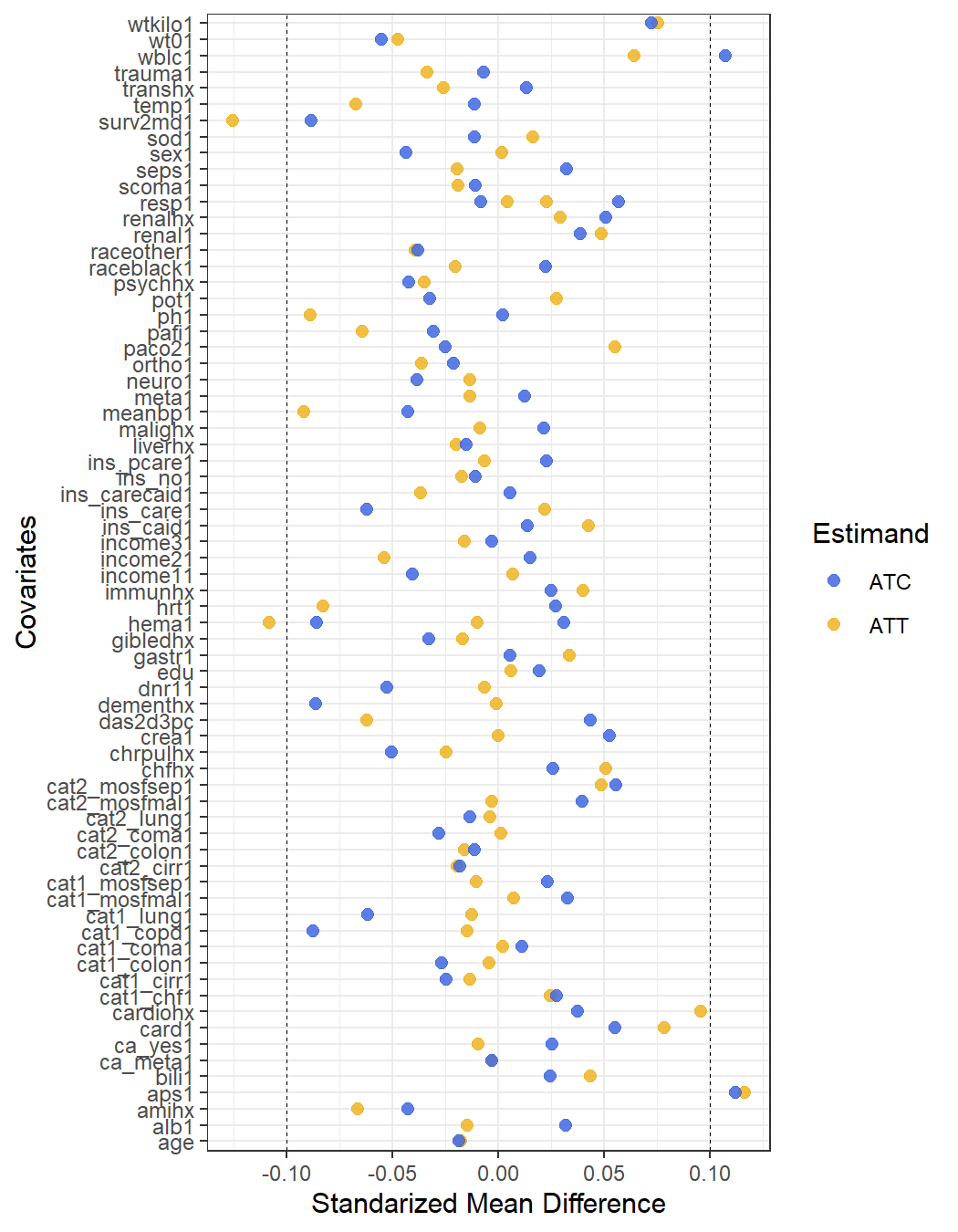}
		\caption{Covariates Balance of RHC Data}\label{Figure_LP.RHC}
	\end{figure}

	Figure \ref{Figure_PS.RHC} shows the propensity score distributions and Figure \ref{Figure_LP.RHC}  gives the standardized difference of covariates between the treatement and control group. The distributions of the estimated propensity scores indicate a good overlap with almost no anticipated extreme weights. Most standardized differences of the covariates are below 0.1 threshold, except for 3 covariates with ATT weights and 2 with ATC weights. The contribution of the 3551  control (resp. 2184 treated)  participants  in effectively estimating the ATT (resp. ATC) via the weights was of about 16\%  (resp. 28.44\%), as highlighted in Table \ref{tab_rhc-ess}.
	
		\begin{table}\small
		\begin{threeparttable}
			\caption {Actual and Effective Sample Sizes (ESS) of the RHC Cohort}\label{tab_rhc-ess}
			\centering
			\begin{tabular}{rcccccc}
				\midrule
				& & & \multicolumn{4}{c}{Effective Sample Size }\\
				\cmidrule(lr){4-7}
				Treatment & N  & \%N & ATT & \%ESS & ATC & \%ESS\\
				\hline
				No RHC  & 3551 & 62 & 567.38 & 15.98 &3551 & 100 \\ 
				RHC  & 2184 & 38 & 2184 & 100 & 621.17 & 28.44 \\ 
				\bottomrule
			\end{tabular}
			\begin{tablenotes}
				\tiny
				\item ATT (resp. ATC): average treatment effect on the treated (resp. controls).  
			\end{tablenotes}
		\end{threeparttable}
	\end{table} 

	\begin{table}[h]\small
		\begin{threeparttable}
		\centering
		\caption{Causal effects and variance estimations from the SUPPORT data on the use of RHC}\label{table_RHC}
		\begin{tabular}{llccr}
			\toprule
			Estimand & Method & Estimate & SE & p-value \\ 
			\midrule
			\multirow{5}{*}{ATT}  
			& Sand. & 0.10 & 0.043 & 0.021 \\ 		
			& WB R & 0.10 & 0.053 & 0.060 \\
			& WB E & 0.10 & 0.046 & 0.032 \\ 
			& Std.Boot & 0.10 & 0.045 & 0.029 \\
			& TMLE & 0.06 & 0.048 & 0.19 \\
			\addlinespace
			\multirow{5}{*}{ATC} 
			& Sand. & 0.15 & 0.037 & $<0.001$ \\
			& WB R & 0.15 & 0.033 & $<0.001$ \\ 
			& WB E & 0.15 & 0.032 & $<0.001$ \\ 
			& Std.Boot & 0.15 & 0.047 & 0.002 \\
			& TMLE & 0.21 & 0.034 & $<0.001$ \\
			\bottomrule
		\end{tabular}
	\begin{tablenotes}
		\tiny
		\item  Sand.: sandwich; WB R (resp. E) : wild bootstrap via Rademacher (resp. exponential) distribution; Std.Boot: standarded bootstrap; TMLE: targeted maximum likelihood estimation.
	\end{tablenotes}
    \end{threeparttable}
	\end{table}
	
	{\color{black}We show, in Table \ref{table_RHC}, the estimated average treatment effect on the treated (ATT) (resp. on controls (ATC)) of the use of RHC on the log length of stay in ICU. Among the treated participants, the effect is equal to 0.10 using the doubly robust estimator \eqref{eq:dr_est}  and 0.06 by TMLE. There are some differences in the estimated standard errors across the different methods. The sandwich variance estimation is the most efficient method (SE = 0.043) while the TMLE (SE = 0.048) and the  wild bootstrap with  Rademacher distribution (SE = 0.053) are the least efficient. The ATT effect of RHC on the log length of stay was significant, except for the wild bootstrap with  Rademacher distribution (p-value = 0.06) and TMLE  (p-value = 0.19).
		
	The average treatment effect on the control	(ATC) participants is statistically significant, regardless of the method. Nevertheless, there is also a discrepancy between the point estimate from the  doubly robust estimator \eqref{eq:dr_est}, which is equal to 0.15,  and that of the TMLE (i.e., 0.21). Furthermore, the wild bootstrap methods are the most efficient with SE = 0.32 and 0.33, respectively,  based on  exponential and Rademacher distributions. The least efficient methods are the sandwich variance estimation (SE = 0.037) and the standard bootstrap (SE = 0.047).}
	
	\section{Conclusion}\label{sec:conclusion}
	{\color{black}In this paper, we propose 3 methods to estimate the variance of the doubly robust ATT and ATC that incorporate the uncertainties inherent to the estimation of the propensity score and regression models. These methods are large-sample sandwich variance estimator and two wild bootstrap methods, which were compared to the standard bootstrap and targeted maximum likelihood estimation (TMLE) methods. The version of TMLE we considered did not include any iterative algorithm this method is famous for. While the sandwich variance estimation is based on asymptotic normal distribution and Slusky's thereom (via M-theory) to yield an efficient, doubly-robust estimators for ATT and ATC when all the models are correctly specified, the wild bootstrap uses the efficient influence function of the estimands that is also guaranteed to lead to efficient variance estimates. Unlike with standard bootstrap where we resample with repetition from the original data set, with wild bootstrap algorithm we perturb the efficient influence function of the underlying estimand using independent and identically distributed random variables to derive the desired variance estimates.
	
	We have provided the underlying theoretical underpinnings of the sandwich variance estimation, examples how the related matrices can be calculated when a logistic regression model is used to estimate the propensity scores and linear regression regression models are considered for the regression models. When the propensity score model is estimated by other regression models than the logistic regression or even through machine learning methods (see Zhou et al. for a list of possible methods \cite{zhou2020propensity}), deriving the matrices needed to estimate the sandwich variance  can be intractable. In that case, the wild  or even just the standard bootstrap methods can constitute reasonable alternatives. The algorithm of the standard bootstrap is simple and commonly known in research circles.  Moreover, we also outlined the algorithm for the wild bootstrap methods. Our finite-sample simulation results indicate that, overall, the sandwich variance estimation yields a reasonable performance, but the bootstrap methods (standard and wild) performed better in good number of scenarios and in particular when the sample size is small. The TMLE method did not performed to the best of its ability probably due to the simplify version we chose. All the methods were less prone to bias and yielded smaller variances when at least the regression models were correctly specified than when the propensity score model alone was correctly specified. Our data application yields similar results across the different methods considered. We applied the variance estimation methods to  the right heart catheterization (RHC) data.
	
	In our assessment of the finite-sample properties of the different methods, appropriate model specifications and good overlap were important factors in providing better variance estimators and the performance of the methods also depended on  whether the treatment was constant or not and whether we estimated ATT or ATC.  Model specification should be done with careful consideration as recommended elsewhere.\cite{zhou2020propensity} When estimating the propensity scores, for instance, we should only include in the model variables that are confounding (i.e., related to both treatment assignment and the outcome) and those that are prognostic (solely related to the outcome). Best practice recommends that the choice of variables to include in any model should be rooted in and built on empirical evidence, related to the scientific question(s) of interest, and take into account elicited expert-matter knowledge of the field of investigation. We strongly caution against blind use of overfitting models, i.e., kitchen-sink type approach where all the variables available in the data set are included in the model, along with all possible two-by-two and even three terms interactions. We do not also recommend backward, forward  and stepwise selection of variables if the goal is to estimate causal effects. Such procedures do not take into account proper causal pathways; they select parsimonious models by relying on a number of inherent test statistics that are not statistically sound (multiple testing) to drop (or keep) variables in the final  model. Moreover, the method often lead to choice of variables that are not causally inclined (inclusion of mediators or colliders) or sometimes not even clinically reasonable or relevant.\cite{harrell2017regression} Finally, it should be also noted that the use of measures of model fit or prediction such as the Hosmer-Lemeshow goodness-of-test, area under the ROC curve, or Akaike's Information Criterion is ill-advised  and irrelevant to select the sought-after models. Sometimes it helps to carefully draw a directed acyclic graph to visually grasp the underlying structure of the data and aid to identify variables that  shouldn't enter a model (e.g., instrumental variables, mediators, and colliders).\cite{shrier2008reducing}   
	
	In the presence of poor overlap and extreme weights, all the methods investigated are expected to perform poorly, despite the slight differences in inference we may notice.  While the influence of poor overlap and extreme weights exert on estimators of ATT and ATC are less preeminent and less frequent than when we estimate the ATE, it is nonetheless present. To estimate ATT we need to monitor control participants that have a propensity score near 1, while for estimating ATC we investigate treated participants with a propensity score near 0. In both scenarios, when extreme weights are present, the risk of getting biased, inefficient, and spurious results is elevated. The current literature recommends to truncate or trim extreme weights \cite{cole2008constructing}. However, most threshold choices  are ad hoc and subjective. Thus, the methods often lead to biased and sometimes inefficient estimators. Moreover, the related estimators do no longer target   the original estimand, but a modified version of it. \cite{ma2020robust, crump2006moving} Further research is needed to provide threshold-independent estimators similar to the overlap weight estimators.\cite{li2018addressing}   
	
	While the wild bootstrap methods to estimate the variance have been investigated proficiently in propensity score matching, \cite{bodory2020finite} this is the first paper to use such methods in propensity score weighting. Obviously, the wild bootstrap methods are also valid  and can  be used to estimate the variance of the average treatment effect (ATE) since its efficient influence function is pretty known.\cite{hirano2003efficient} Similar methods can be used to estimate the variance  of the nascent estimators of ATE-like estimands when lack of positivity is expected such as the overlap weights, the matching weights, the entropy weights, and the beta weights.\cite{matsouaka2020framework}  Nevertheless, further investigations are warranted to assess their final-sample performance to that of the standard bootstrap, the sandwich variance, and the TMLE variance estimation methods for these estimand, including the classic ATE. Finally, to conform with our data generating processes and to ensure comparability, we did not leverage  or expand on the interesting features the TMLE provides.  We encourage the reader to delve into the related literature for an	excellent treatment of the topic.\cite{schuler2017targeted}
	
	Overall, to estimate the variance of the ATT or ATC, one can use the sandwich variance estimation methods as it is robust, conservative, and performs fairly well. When the sandwich variance is not applicable (intractable matrix derivation or small sample size), we recommend  wild bootstrap methods as they are more flexible and robust in estimating the desired variances. When the  sample size is small, we need to use  the standard bootstrap estimation. Nevertheless, we caution that small-sample results are usually unstable; care must be taken in interpreting and presenting such results in a sensible manner.}
	\bibliographystyle{SageV}
	\bibliography{variance_dr_att}
	
	\newpage	
	\appendix
\section{Supplemental Material: Technical proofs}\label{appendix}
 \setcounter{equation}{1}

\addtocounter{equation}{1}

\newcounter{Appendix}[section]
\numberwithin{equation}{subsection}
\renewcommand\theequation{\Alph{section}.\arabic{subsection}.\arabic{equation}}
\numberwithin{table}{subsection}
\numberwithin{figure}{subsection}
\subsection{Sandwich variance estimation}\label{sec:sandvariance_estimation}

In this section we present variance estimation of the ATT and ATC using M-theory (see Stefanski and Boos \cite{stefanski2002calculus}). The variance of an  estimator $\widehat\tau_{g}=c'\theta$, for some vector $c$, is derived using an estimating equation of the form $\displaystyle \displaystyle\sum_{i=1}^{N} \Psi_\theta({X}_i, Z_i, Y_i)=0$ for which $\theta$ is the solution, for some matrix $ \Psi_\theta({X}_i, Z_i, Y_i).$ 

Starting with $E[\Psi_\theta({X}_i, Z_i, Y_i)]=0$, we can show that $\widehat\theta\overset{p}{\longrightarrow}  \theta $ when  $N\longrightarrow \infty$, under some  regularity conditions.  \cite{stefanski2002calculus} Thus, by Slutsky's theorem, $ \widehat\theta$ is consistent for $\theta$. In addition,  $\sqrt{N}(\widehat{\theta}-{\theta})\overset{d}{\longrightarrow}N({0}, \Sigma({\theta})),$ with $ \Sigma({\theta})=A(\theta)^{-1}B(\theta)\{A(\theta)'\}^{-1}$. A consistent estimator of variance-covariance matrix $\Sigma({\theta})$ of $\widehat{\theta}$ is  $ \widehat \Sigma({\widehat\theta})=A_N(\widehat\theta)^{-1}B_N(\widehat\theta)\{A_N(\widehat\theta)'\}^{-1}$, where  $A(\theta), B(\theta),$ $ A_N(\widehat\theta)$ and $B_N(\widehat\theta)$ are the following matrices: 
\begin{align*}
	A_N(\widehat{\theta})&=\frac{- 1}{N}\displaystyle\sum_{i=1}^{N}\frac{\partial \Psi_\theta({X}_i, Z_i, Y_i)}{\partial{\theta'}}\Big|_{\theta={\widehat\theta}}; ~~B_N(\widehat{\theta})= \frac{1}{N}\displaystyle\sum_{i=1}^{N}\Psi_\theta({X}_i, Z_i, Y_i)\Psi_\theta({X}_i, Z_i, Y_i)'\big|_{\theta=\widehat\theta};
	\\
	A(\theta)&=\lim_{N\rightarrow\infty}A_N(\widehat\theta)  ~\text{and}~~
	B(\theta)=\lim_{N\rightarrow\infty}B_N(\widehat\theta)=E\left[ \Psi_\theta({X}_i, Z_i, Y_i)\Psi_\theta({X}_i, Z_i, Y_i)'\right]. 
\end{align*} 
The estimator of the variance of $\widehat\tau_{g}=c'\theta$ is then $ c'\widehat \Sigma({\widehat\theta})c.$

As an illustrative example, we provide the matrices $A_N$ and $B_N$ when the propensity score model $e({x_i};\widehat{\beta})=P(Z=1|{X_i=x_i}; \widehat{\beta}))$ and the regression models  $\widehat m_z(x_i)=m(x_i; \widehat\alpha_z)$ for $z=0$ and $z=1$ are estimated by maximum likelihood using, respectively, the logistic and linear regression models---where $\widehat{\beta}$ (resp. $\widehat{\alpha}_z$) is an estimator of $\widehat{\beta}$  (resp. ${\alpha}_z$), a vector with $p+ 1$ (resp. $q+1$) components. We consider that different combinations (or subsets) of covariates $X$ enter the logistic and regression models, which we  denote respectively $V$ and $W.$ 

\subsubsection{Variance for the weighted mean estimator }\label{appendix4_sandwich_wt}   
Consider an estimator $\widehat e(x) =e(x;\widehat \beta)$  of $e(X;{\beta})$, $\widehat  g(x)= a + b\widehat e(x)$,  and  $\widehat w_z(x_i) = t(z, Z_i, x_i)\left( \displaystyle\sum_{i=1}^{N}t(z, Z_i, x_i)\right) ^{-1}$, with	
$ t(z, Z, x) = \left( \displaystyle \frac{Z\widehat g(x) }{\widehat e(x)}\right)^z \left( \displaystyle \frac{(1-Z)\widehat g(x) }{1-\widehat e(x)}\right) ^{1-z},$ where $a, b, z \in \{0, 1\}$. Define the estimator $\widehat{\mu}_{zg}= \displaystyle \displaystyle\sum_{i=1}^{N}\widehat w_z(x_i) Y_i$ of ${\mu}_{zg}=\displaystyle{E[g(X)Y(z)]}/{E[g(X)]}$.

The  WATE estimator is $\widehat\tau_{g}= \widehat{\mu}_{1g}-\widehat{\mu}_{0g}$, which can be written as  	
$\widehat{\tau}_g=c_0'{\theta}=\widehat{\mu}_{1g}-\widehat{\mu}_{0g}$, for $c_0=(\underbrace{0,\dots, 0}_{p+1},1,-1)'$ and $\widehat{\theta}=(\widehat{\beta}', \widehat{\mu}_{1g}, ~\widehat{\mu}_{0g})'$. 	
The vector $\widehat{\theta}$ is solution to the estimating equation	$\displaystyle\sum_{i=1}^{N}\Psi_\theta({X}_i, Z_i, Y_i)
= 0$ with respect to  ${\theta}=({\beta}',\mu_{1g}, \mu_{0g})'$, where
\begin{align*}
	\Psi_\theta({X}_i, Z_i, Y_i)
	= 
	\displaystyle \begin{bmatrix}
		\psi_{\beta}({X}_i, Z_i)\\
		\psi_{\mu_{1g}}({X}_i, Z_i, Y_i) \\
		\psi_{\mu_{0g}}({X}_i, Z_i, Y_i) \\
	\end{bmatrix} 
	= \displaystyle  \begin{bmatrix}
		\psi_{\beta}({X}_i, Z_i)\\
		t(1, Z_i, {X}_i)(Y_i-\mu_{1g})\\
		t(0, Z_i, {X}_i)(Y_i-\mu_{0g})\\
	\end{bmatrix}.
\end{align*}
The matrices $A_N(\widehat\theta)$ and $B_N(\widehat\theta)$ are given by
	$		A_N(\widehat{\theta})=N^{-1}\displaystyle\sum_{i=1}^{N}\left[ - \frac{\partial}{\partial{\theta'}}\Psi_\theta({X}_i, Z_i, Y_i)\right]_{\theta={\widehat\theta}}$ and 
	$		B_N(\widehat{\theta})= \frac{1}{N}\displaystyle\sum_{i=1}^{N}\Psi_\theta({X}_i, Z_i, Y_i)\Psi_\theta({X}_i, Z_i, Y_i)'\big|_{\theta=\widehat\theta}$.

If we estimate the propensity scores via a logistic regression model $e(X_i)=[1+\exp(-V_{i}'\beta)]^{-1}$,  then $\psi_{{\beta}}({X}_i, Z_i)=[Z_i-e({V_{i}};{\beta})]V_{i}$, i.e., the corresponding score function. The non-zero components of the matrix $A_N$ are 
\allowdisplaybreaks\begin{align*}
	\widehat A_{11}&=N^{-1}\displaystyle\sum_{i=1}^{N} \widehat e_i(\mathrm{v})(1-\widehat e_i(\mathrm{v}))V_iV_i', ~~\text{with dimension}~~p\times p; \\
	\widehat A_{21}&=-N^{-1}\displaystyle\sum_{i=1}^{N} Z_i\left[\left[\frac{\partial g(V_i)}{\partial{\beta}}\right]_{\beta=\widehat\beta} - (1-\widehat e_i(\mathrm{v}))\widehat{g}(V_i)V_i'\right]\widehat e_i(\mathrm{v})^{-1} \left(Y_i-\widehat\mu_{1g}\right);\\
	\widehat A_{31}&=-N^{-1}\displaystyle\sum_{i=1}^{N} (1-Z_i)\left[\left[\frac{\partial g(V_i)}{\partial{\beta}}\right]_{\beta=\widehat\beta} + \widehat e_i(\mathrm{v})\widehat{g}(V_i)V_i'\right](1-\widehat e_i(\mathrm{v}))^{-1} \left( Y_i-\widehat\mu_{0g}\right); \\
	\widehat A_{22}&=N^{-1}\displaystyle\sum_{i=1}^{N} Z_i\widehat e_i(\mathrm{v})^{-1}\widehat{g}(V_i); ~~
	\widehat A_{33} =N^{-1}\displaystyle\sum_{i=1}^{N}(1- Z_i)(1-\widehat e_i(\mathrm{v}))^{-1}\widehat{g}(V_i).		
\end{align*}  
The dimensions of the components $\widehat A_{kj}$ will depend on $p$, the number of columns  of $V.$

Note that, in particular, $\displaystyle\left[\displaystyle {\partial g(V_i)}/{\partial{\beta}}\right]_{\beta=\widehat\beta}$ is equal to 0 for ATE, equal to $e_i(\mathrm{v})[1-e_i(\mathrm{v})]V_i'$ for ATT and  to  $-e_i(\mathrm{v})[1-e_i(\mathrm{v})]V_i'$ for ATC.
An estimator of the variance of $\widehat{\tau}_g$  is $\widehat{Var}({\widehat{\tau}_g})=N^{-1}c_0'\widehat\Sigma({\widehat\theta})c_0$.\\

\subsubsection{Variance for the doubly robust ATT and ATC }\label{appendix4_sandwich_dr}   
The doubly robust estimators for  ATT and ATC are given, respectively, by 
\allowdisplaybreaks\begin{align}\label{tau_dr}
	\widehat \tau_{\text{ATT}}^{dr}&=\sum_{i=1}^{N}(\widehat  w_1(x_i)-\widehat  w_0(x_i))\left\lbrace Y_i- \widehat m_0(x_i)\right\rbrace; \\
	\widehat \tau_{\text{ATC}}^{dr}&=\sum_{i=1}^{N}(\widehat  w_1(x_i)-\widehat  w_0(x_i))\left\lbrace Y_i- \widehat m_1(x_i)\right\rbrace.\nonumber
\end{align}
In addition to $\psi_{\beta}({X}_i, Z_i)$,   we also consider the score functions $\psi_{\alpha_z}(X)$ for the models $m_z(X)=m_z(X; \alpha_z),$ $z=0,1$. 
For the ATT estimator $\widehat\tau_{ATT}^{dr}$, we  use the  solution  $\widehat\theta_{att}=(\widehat\beta',  \widehat\alpha_0', \widehat\mu_{1}, \widehat\mu_{0})'$ to 
\allowdisplaybreaks \begin{align*}
	\displaystyle \displaystyle\sum_{i=1}^{N} \Psi_{\theta_{att}}({X}_i, Z_i, Y_i)&= 
	\displaystyle \displaystyle\sum_{i=1}^{N}	\begin{bmatrix}
		\psi_{\beta}({X}_i, Z_i)\\
		(1-Z_i)\psi_{\alpha_0}({X}_i, Y_i)\\
		Z_i(Y_i-m_0(X_i)-\mu_{1})\\
		(1-Z_i)e(x_i)(1-e(x_i))^{-1}(Y_i-m_0(X_i)-\mu_{0})\\
	\end{bmatrix} =0,
\end{align*}
with respect to $\theta_{att}=(\beta',\alpha_0',  \mu_{1}, \mu_{0})',$ to calculate       
$\widehat\tau_{ATT}^{dr}=c_2'\widehat\theta_{att}=\widehat\mu_{1}-\widehat\mu_{0}$, where  $c=(\underbrace{0,\dots, 0}_{p+1},\underbrace{0,\dots, 0}_{q+1},1, -1)$ and 
$\widehat{\mu}_{z}= \displaystyle \displaystyle\sum_{i=1}^{N} \widehat w_z(x)\left( Y_i-\widehat m_0(x_i)\right),$  with $z=0, 1.$

When  $e(X)$ and  $m_z(X)$ are estimated via maximum likelihood based on logistic and linear regression models  $e(V_i)=[1+\exp(-V_i'\beta)]^{-1}$ and $m_z(W_i)=W_i'\alpha_z$, $z=0,1$, then $\psi_{{\beta}}({X}_i, Z_i)=[Z_i-e({V_i};{\beta})]V_i$ and $\psi_{{\alpha}_z}({X}_i, Z_i)=W_i(Y_i-W_i'{\alpha}_z)$. 
Assuming that the same covariates appear as predictors in the regression models $m_z(W)$, the non-zero components $\widehat A_{ij}$ of the matrix $ A_N$ are given by 
\allowdisplaybreaks\begin{align*}
	\widehat A_{11}&=N^{-1}\displaystyle\sum_{i=1}^{N} \widehat e_i(\mathrm{v})(1-\widehat e_i(\mathrm{v}))V_iV_i'; ~
	\widehat A_{22}=N^{-1}\displaystyle\sum_{i=1}^{N}(1- Z_i)W_iW_i';\\ 
	\widehat A_{32}&=N^{-1}\displaystyle\sum_{i=1}^{N} Z_iW_i'; ~\widehat A_{33}=N^{-1}\displaystyle\sum_{i=1}^{N} Z_i;\\
	\widehat A_{41}&=-N^{-1}\displaystyle\sum_{i=1}^{N} (1-Z_i)\frac{\widehat e_i(\mathrm{v})}{(1-\widehat e_i(\mathrm{v}))}(Y_i-\widehat m_0(W_i)-\widehat \mu_{0})V_i'; \\
	\widehat A_{42}&=N^{-1}\displaystyle\sum_{i=1}^{N}(1- Z_i) \frac{\widehat e_i(\mathrm{v})W_i'}{(1-\widehat e_i(\mathrm{v}))};~~
	\widehat A_{44}=N^{-1}\displaystyle\sum_{i=1}^{N}(1- Z_i) \frac{\widehat e_i(\mathrm{v})}{(1-\widehat e_i(\mathrm{v}))}.
\end{align*}   
The variance of  $\widehat\tau_{\text{ATT}}^{dr}$ is then estimated through $\widehat{Var}(\widehat{\tau}_{\text{ATT}}^{dr})=N^{-1}c'\widehat\Sigma({\widehat{\theta}_{att}})c.$

With ATC, the estimator        
$\widehat\tau_{ATC}^{dr}=c'\widehat\theta_{atc}=\widehat\mu_{1}-\widehat\mu_{0}$ where $\widehat\theta_{atc}=(\widehat\beta', \widehat\alpha_1',  \widehat\mu_{1}, \widehat\mu_{0})'$ is the solution   to the estimating equation
\allowdisplaybreaks \begin{align*}
	\displaystyle \displaystyle\sum_{i=1}^{N} \Psi_{\theta_{atc}}({X}_i, Z_i, Y_i)&= 
	\displaystyle \displaystyle\sum_{i=1}^{N}		\begin{bmatrix}
		\psi_{\beta}({X}_i, Z_i)\\
		Z_i\psi_{\alpha_1}({X}_i, Y_i)\\
		Z_ie(x_i)^{-1}(1-e(x_i))(Y_i-m_1(X_i)-\mu_{1})\\
		(1-Z_i)(Y_i-m_1(X_i)-\mu_{0})\\
	\end{bmatrix} =0
\end{align*}
with respect to  $\theta_{atc}= (\beta',\alpha_1',   \mu_{1}, \mu_{0})'$  and
$\widehat{\mu}_{z}= \displaystyle \displaystyle\sum_{i=1}^{N}\widehat  w_z(x_i)\left( Y_i-\widehat m_1(x_i)\right)$,  for $z=0, 1.$

In this case, the non-zero components of the matrix $A_N$ are then 
\allowdisplaybreaks\begin{align*}
	\widehat A_{11}&=N^{-1}\displaystyle\sum_{i=1}^{N} \widehat e_i(\mathrm{v})(1-\widehat e_i(\mathrm{v}))V_iV_i'; ~
	\widehat A_{22}=N^{-1}\displaystyle\sum_{i=1}^{N} Z_iW_iW_i';\\ 
	\widehat A_{31}&=N^{-1}\displaystyle\sum_{i=1}^{N} \frac{Z_i(1-\widehat e_i(\mathrm{v}))}{\widehat e_i(\mathrm{v})}(Y_i-\widehat m_1(W_i)-\widehat \mu_{1})V_i'; 	\\
	\widehat A_{32}&=N^{-1}\displaystyle\sum_{i=1}^{N} \frac{Z_i(1-\widehat e_i(\mathrm{v}))}{\widehat e_i(\mathrm{v})}W_i'; ~~
	\widehat A_{33}=N^{-1}\displaystyle\sum_{i=1}^{N} \frac{Z_i(1-\widehat e_i(\mathrm{v}))}{\widehat e_i(\mathrm{v})}; \\
	\widehat A_{42}&=N^{-1}\displaystyle\sum_{i=1}^{N} (1-Z)W_i';~~
	\widehat A_{44}=N^{-1}\displaystyle\sum_{i=1}^{N}(1- Z_i).
\end{align*}   
An estimator of the variance of  $\widehat\tau_{{ATC}}^{dr}$ is given by $\widehat{Var}(\widehat{\tau}_{{ATC}}^{dr})=N^{-1}c'\widehat\Sigma({\widehat{\theta}_{atc}})c.$

\section{Supplemental Material: Simulation Results}\label{appendix2}
\subsection{Propensity score analysis}\label{appendix2_PS}

\begin{figure}[h]
	\centering
	\includegraphics[width=1\textwidth]{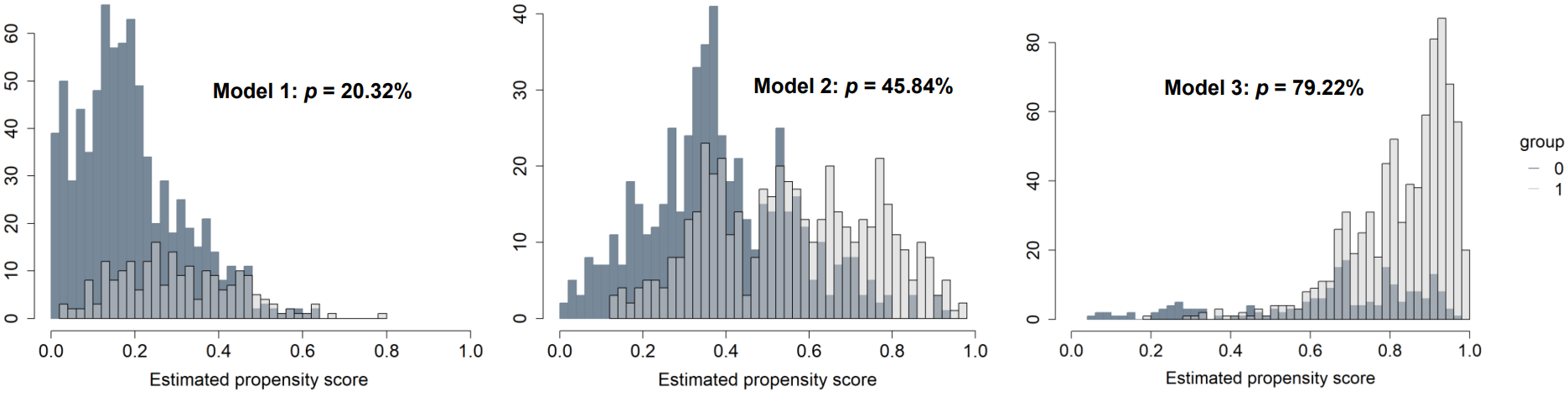}
	\caption{Estimated propensity scores for the different models simulating different proportions of treated participants (model 1, 2 and 3)}\label{fig:ps_md123}
\end{figure}

\begin{figure}[h]
\centering
\includegraphics[width=0.55\textwidth]{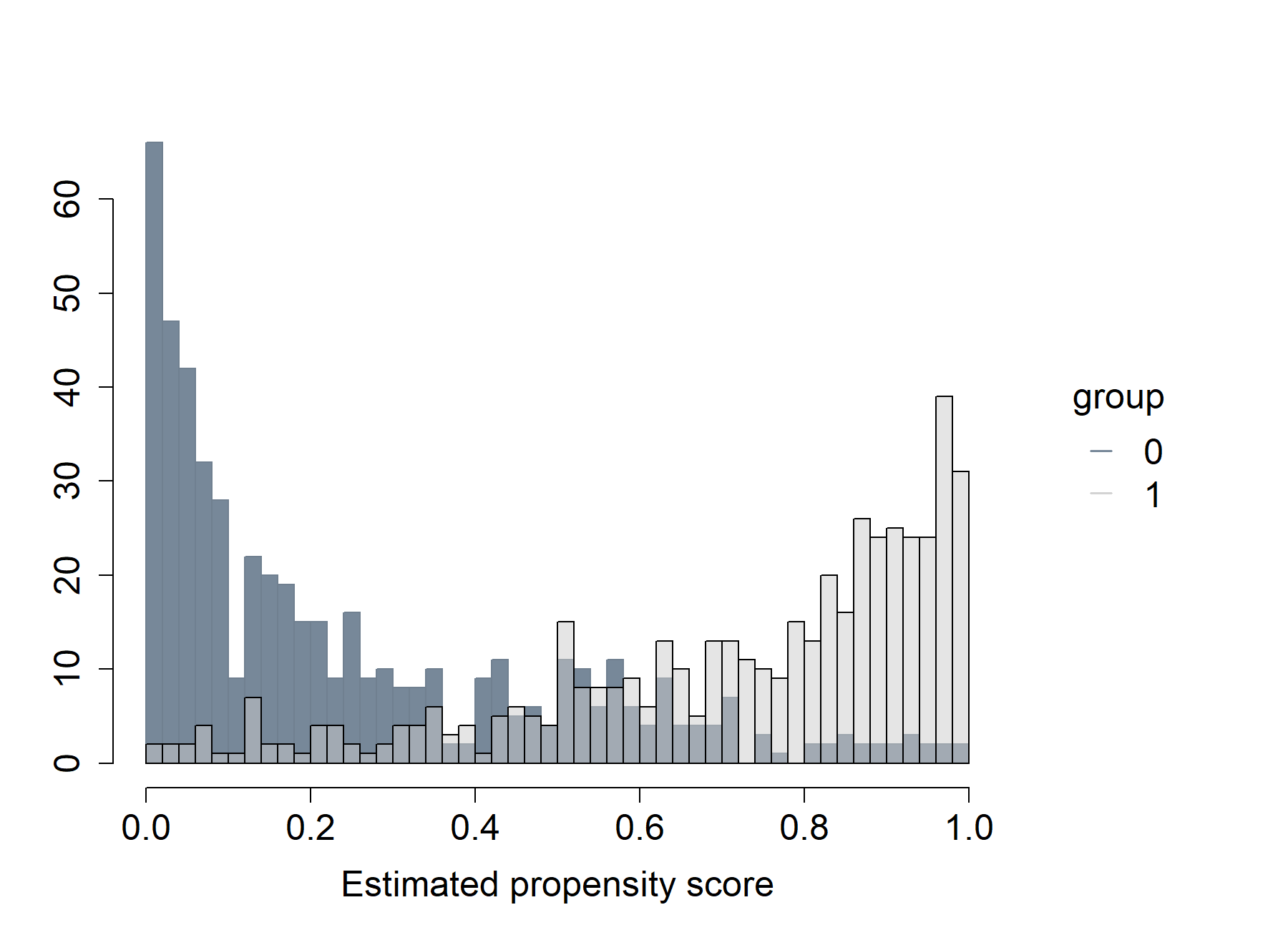}
\caption{Estimated propensity scores for the case of poor overlap and extreme weights exist (model 4)}\label{fig:ps_md4}
\end{figure}

\begin{figure}[h]
\centering
\includegraphics[width=1\textwidth]{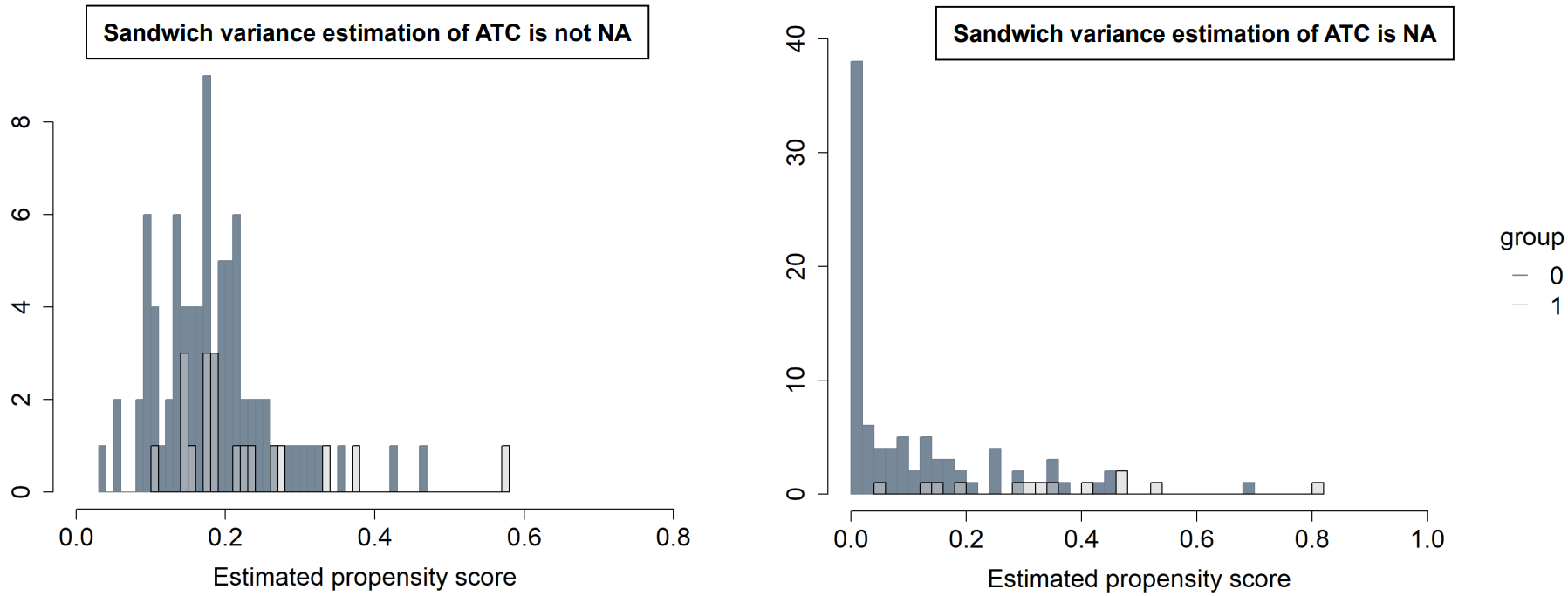}
\caption{Estimated propensity scores for the case when sandwich variance estimation does not always work (model 5a, with $N=100$): the left figure is by a replication where the sandwich variance estimation for ATC is not NA, where the right figure is by a replication where the sandwich estimation for ATC is NA.}\label{fig:ps_md5}
\end{figure}

\color{black}
As indicated in Section \ref{sec:simulations}, we considered 5 different propensity score (PS) models as well as both homogeneous and heterogeneous treatment effect for each model setting. For each correctly specified PS model, a glimpse of  related propensity score distributions is given (see Figure \ref{fig:ps_md123} to \ref{fig:ps_md5}), from a randomly selected data replicate. Furthermore, by taking average over 2000 replicates, we provide the effective sample sizes (ESS) in Table \ref{tab_ESS}, which reflect the performance of weights under different proportions. 

\begin{table}[h]\small
	\begin{threeparttable}
		\caption {Effective Sample Sizes (ESS)}\label{tab_ESS}
		\centering
		\begin{tabular}{ccccccccccccccc}
			\toprule
			& \multicolumn{2}{c}{\textbf{Model 1}} & \multicolumn{2}{c}{\textbf{Model 2}} & \multicolumn{2}{c}{\textbf{Model 3}} & \multicolumn{2}{c}{\textbf{Model 4}} & \multicolumn{2}{c}{\textbf{Model 5}} \\
			\cmidrule(lr){2-3}\cmidrule(lr){4-5}\cmidrule(lr){6-7}\cmidrule(lr){8-9}\cmidrule(lr){10-11}\cmidrule(lr){12-13}
			Treatment  & ATC & ATT & ATC & ATT & ATC & ATT & ATC & ATT & ATC & ATT \\ 
			\midrule
			0  & 796.82 & 382.34 & 541.61 & 268.62 & 207.78 & 98.17 & 510.38 &  53.32 & 79.75 & 31.79 \\ 
			1  & 93.92 & 203.18 & 206.13 & 458.39 & 306.68 & 792.22 & 51.78 & 489.62 & 9.30 & 20.25 \\ 
			\bottomrule
		\end{tabular}
		\begin{tablenotes}
			\tiny
			\item ATT (resp. ATC): average treatment effect on the treated (resp. controls)
			\item Note that for model 5, $N=100$. 
		\end{tablenotes}
	\end{threeparttable}
\end{table} 

The purpose of designing models 1--3 is to investigate the performance of different methods under different proportions $p=P(Z=1)$ of participants who receive  treatment. We defined  $Z\sim \text{Bern}\left(  (1+\exp(-X\beta))^{-1}\right)$, where   $\beta=(\beta_0, 0.3, 0.4, 0.4, 0.4, -0.1, -0.1, 0.1)'$ such that 
\begin{itemize}[topsep=0pt, noitemsep]
	\item Model 1: $\beta_0 =-2.17, p = 20.32\%$;
	\item Model 2: $\beta_0 =-0.78, p = 45.84\%$;
	\item Model 3: $\beta_0 =0.98, p = 79.22\%$.
\end{itemize}
Figure \ref{fig:ps_md123} gives the distributions of the estimated propensity scores of these 3 models. The corresponding results are given in Tables \ref{sim-model1.effect}--\ref{sim-model5.effect}.

For	Model 4, we chose $\beta = (0.2, 1.0, -0.9, -0.9, 0.9, 0.15, 0.15, -0.2)'$ for which {$p=46.65\%$}. This model represents the scenario with poor overlap and extreme weights (see Figure \ref{fig:ps_md4}). 
Finally, for the last scenario, we considered Model 1 but under small sample sizes  ($N=100$ and $N= 50$), which we labeled "Model 5a" and "Model 5b", respectively. These models pertain to  situations where the sandwich variance estimator may not be applicable since the sandwich variance method is based on  large sample approximations. 	We provide the true heterogeneous treatment effects by model in Table \ref{tab:truth}. Based on our DGP mentioned in Section \ref{sec:simulations}, the true constant effects are 4 for all cases and models.

\begin{table}[h]\small\color{black}
\begin{threeparttable}
	\caption {True heterogeneous treatment effects by model}\label{tab:truth}
	\centering
	\begin{tabular}{crrr}
		\toprule
		Model & Case & ATT & ATC \\ 
		\midrule 
		1 & small $p$ & 20.93 & 16.27  \\ 
		2 & medium $p$ & 18.34 & 16.26 \\ 
		3 & large $p$ & 16.86 & 18.58 \\ 
		4 & poor overlap and extreme weights exist (with medium $p$) & 18.67 & 15.83 \\ 
		5a & sandwich variance does not work ($N=100$) & 20.93 & 16.27 \\ 
		5b & sandwich variance does not work ($N=50$) & 20.93 & 16.27 \\ 
		\bottomrule
	\end{tabular}
	\begin{tablenotes}
		\tiny
		\item $p=P(Z=1)$ is the proportion of the treated participants
	\end{tablenotes}
\end{threeparttable}
\end{table}

As expected, we there was a number of data replicates where the sandwich did not work and failed to return valid variance estimates.
We report, in Table \ref{tab:NAsmaller}, the frequency of cases where the sandwich variance estimate was unobtainable
over 2000 data replicates under different model specifications. Interestingly, the results (i.e., how many unobtainable sandwich variance estimate 
we had) were exactly the same for both constant and heterogeneous treatments.  On the one hand, there was no empty variance estimate 
when estimating sandwich variances for ATT for $N=100$ and just a handful when $N=50$. On the other hand, estimations of the sandwich variance for ATC were sensitive to the size of the data. Evidently, the smaller the size of the data, the higher the number of unobtainable sandwich variance estimate
in sandwich variance estimations. Overall, for ATC, there were about 1\% of data replicates that fail to provide the sandwich variance estimate for $N=100$ whereas for $N=50,$ the percentage of nonobtainable sandwich variance estimates
varied between 20\% and 36\%, depending on the specifications of the parametric models we used  (see Table \ref{tab:NAsmaller}).   

The fact we can estimate the sandwich variance relatively well for ATT even for $N=50$ but not quite well for ATC
indicates that the true issue is related with the presence or not of extreme propensity score weights. For ATT, we did not expect to see extreme weights  since  $\widehat w_1(x)=1/N$ and all $e_i(x)$, for control participants are far away from 1, i.e., they cannot lead to large values of $\widehat w_0(x)$. On the opposite, for ATC a number of $e_i(x)$'s would be very small for treated participants and their variability extremely unstable. This often happens when the proportion of treated participants is also small ($p=20.32\%$ in Model 5a and 5b). Thus, the  combination of small sample size and  small $p$ may be particularly damaging to sandwich variance estimation of ATC.

Figure \ref{fig:ps_md5} gives the estimated propensity scores of two typical data replicates from this scenario, i.e., one when we can obtain the sandwich variance estimation (of ATC) and the other when the estimation leads to an unobtainable sandwich variance estimate, 
for $N=100$. This simple exercise gives an elegant demonstration of  what happens when small sample size collude with small proportion of participants in one of the treatment groups and how unstable sandwich variance estimation can be. When models are too complex vis-\`a-vis the amount of data (and information) available, the performance of the related logistic regression models is often unstable and less reliable. It can be seen that when sandwich variance estimation (of ATC) is not obtainable, almost all propensity scores of the control group are in the range of $[0,0.02]$ (very small), which might indicate that these samples have high collinearity, which makes the matrix $\widehat A(\theta)$ singular (see Appendix \ref{appendix}, Section \ref{appendix4_sandwich_dr}). Thus the sandwich variance estimation cannot be obtained. The smaller the sample size, the more likely we encounter this ``high collinearity'' situation.

\begin{table}[h]\small\color{black}
\begin{threeparttable}
	\caption {Frequency of nonobtainable sandwich variance estimate from Model 5, by sample size.}
	\label{tab:NAsmaller}
	\centering
	\begin{tabular}{cccc}
		\toprule
		& & \multicolumn{2}{c}{number of nonobtainable variance estimates}\\ \cmidrule(lr){3-4}
		Estimand  & Correctly specified model(s)		& $N=50$& $N=100$\\ 
		\midrule 
		\multirow{4}{*}{ATC} 
		& PS and OR models & 720 & 19\\ 
		& PS  model  & 403 & 16 \\ 
		& OR models  & 719 & 19 \\ 
		& None & 400 & 16 \\ 
		\midrule
		\multirow{4}{*}{ATT} 
		& PS and OR models & 8 & 0 \\ 
		& PS  model   & 6 & 0 \\ 
		& OR models & 0 & 0 \\ 
		& None   & 0 & 0 \\ 
		\bottomrule
	\end{tabular}
	\begin{tablenotes}
		\tiny
		\item Number of nonobtainable variance estimates obtained  over 2000 replicates
	\end{tablenotes}
\end{threeparttable}
\end{table}

\color{black}	

\subsection{Variance estimations and summary statistics}\label{appendix2_VE}

This section gives the variance estimations and comparison measures for Models 1--4 and Model 5a mentioned in Section \ref{appendix2_PS}. 

{\color{black} From the results by Models 1 to 4, where the sample size is large ($N=1000$), we found that the two wild bootstrap methods perform overall the best in regard to coverage probability (CP). For Model 5a, when the sample size is small ($N=100$), the standard bootstrap overall has the best CP, where sandwich variance estimation and two wild bootstrap methods sometimes result in CPs that lower a lot than 0.95, and they seem to overestimate the efficiency of the doubly robust estimations of ATT and ATC. These might reflect that the performance of wild bootstrap and sandwich variance estimation rely on large sample sizes, because both methods depend on the asymptotic behavior of the semiparametric estimator.}

\begin{table}\scriptsize
\begin{threeparttable}
	\centering
	\caption{Model 1, with $p = 20.32\%$}
	\label{sim-model1.effect}
	\begin{tabular}{llcccccccccccc}
		\toprule	
		& & \multicolumn{12}{c}{Constant treatment effect} \\\cmidrule(lr){3-14}
		& &\multicolumn{6}{c}{PS and OR models correctly specified} & \multicolumn{6}{c}{PS model correctly specified} \\\cmidrule(lr){3-8}\cmidrule(lr){9-14} 
		Est. & Method & Bias & RMSE & SE & ESD & RE & CP & Bias & RMSE & SE & ESD & RE & CP \\ 
		\midrule
		\multirow{5}{*}{ATT} 
		& Sand &  0.04 & 0.18  & 0.33 & 0.18 & 0.29 & 1.00 & 2.88 & 0.96 &  0.87 & 0.95 & 1.19 &  0.99 \\ 			
		& WB R &  0.04 & 0.18 & 0.17 & 0.18 & 1.05 & 0.95 & 2.88 & 0.96 &  1.88 & 0.95 & 0.25 & 1.00 \\ 
		& WB E  &  0.04 & 0.18 & 0.17 & 0.18 & 1.06 &  0.95 & 2.88 & 0.96 &  1.81 & 0.95 & 0.28 & 1.00 \\
		& Std.Boot & 0.04 & 0.18 & 0.17 & 0.18 & 1.02 & 0.95 & 2.88 & 0.96 & 0.78 & 0.95 & 1.48 & 0.97 \\ 
		& TMLE & 2.40 & 0.38 & 0.30 & 0.37 & 1.51 & 0.94 & 4.29 & 1.03 & 1.90 & 1.01 & 0.28 & 1.00 \\ 
		\addlinespace
		\multirow{5}{*}{ATC} 
		& Sand &  0.06 & 0.23  & 0.25 & 0.23 & 0.85 & 0.97 & 19.42 & 4.30 & 2.57 & 4.23 & 2.70 &  0.89 \\  
		& WB R & 0.06 & 0.23 & 0.20 & 0.23 & 1.34 & 0.93 & 19.42 & 4.30 & 2.95 & 4.23 & 2.06 & 0.90 \\ 
		& WB E & 0.06 & 0.23 & 0.19 & 0.23 & 1.44 & 0.91 & 19.42 & 4.30 & 2.49 & 4.23 & 2.88 & 0.87 \\
		& Std.Boot & 0.06 & 0.23 & 0.23 & 0.23 & 1.05 & 0.95 & 19.42 & 4.30 & 1.98 & 4.23 & 4.57 & 0.78 \\ 
		& TMLE & 21.05 & 1.65 & 1.01 & 1.42 & 1.98 & 0.74 & 10.37 & 1.79 & 1.58 & 1.74 & 1.22 & 0.91 \\
		\addlinespace 
		& &\multicolumn{6}{c}{PS and OR models misspecified} & \multicolumn{6}{c}{OR model correctly specified} \\\cmidrule(lr){3-8}\cmidrule(lr){9-14} 
		&  & Bias & RMSE & SE & ESD & RE & CP & Bias & RMSE & SE & ESD & RE & CP \\ 
		\midrule
		\multirow{5}{*}{ATT}  
		& Sand. & 33.95 & 2.23  & 1.66 & 1.76 & 1.13 & 0.88 & 0.03 & 0.17 &  0.33 & 0.17 & 0.28 & 1.00 \\ 
		& WB R  & 33.95 & 2.23 & 2.01 & 1.76 & 0.77 & 0.96 & 0.03 & 0.17 &  0.17 & 0.17 & 1.08 &  0.95 \\ 
		& WB E  & 33.95 & 2.23 & 1.93 & 1.76 & 0.84 & 0.94 & 0.03 & 0.17 & 0.17 & 0.17 & 1.09 &  0.94 \\ 
		& Std.Boot & 33.95 & 2.23 & 1.60 & 1.76 & 1.21 & 0.87 & 0.03 & 0.17 & 0.17 & 0.17 & 1.02 & 0.95 \\ 
		& TMLE & 31.94 & 2.24 & 1.96 & 1.84 & 0.88 & 0.94 & 1.19 & 0.21 & 0.21 & 0.21 & 0.95 & 0.96 \\
		\addlinespace 
		\multirow{5}{*}{ATC}
		& Sand. &  4.58 & 4.14 & 2.69 & 4.14 & 2.36 & 0.86 & 0.09 & 0.22 & 0.25 & 0.22 & 0.76 & 0.97 \\ 
		& WB R & 4.58 & 4.14 &  3.51 & 4.14 & 1.39 & 0.89 & 0.09 & 0.22 &  0.21 & 0.22 & 1.10 &  0.94 \\ 			  
		& WB E & 4.58 & 4.14 & 2.91 & 4.14 & 2.02 & 0.87 & 0.09 & 0.22 &  0.20 & 0.22 & 1.20 &  0.93 \\
		& Std.Boot & 4.58 & 4.14 & 2.64 & 4.14 & 2.46 & 0.85 & 0.09 & 0.22 & 0.22 & 0.22 & 0.96 & 0.95 \\ 
		& TMLE & 14.52 & 2.47 & 2.00 & 2.40 & 1.45 & 0.84 & 13.01 & 1.16 & 1.02 & 1.03 & 1.03 & 0.93 \\ 
		\addlinespace 
		& &\multicolumn{12}{c}{Heterogeneous treatment effect} \\\cmidrule(lr){3-14}
		& &\multicolumn{6}{c}{PS and OR models correctly specified} & \multicolumn{6}{c}{PS model correctly specified} \\\cmidrule(lr){3-8}\cmidrule(lr){9-14} 
		& Method & Bias & RMSE & SE & ESD & RE & CP &  Bias & RMSE & SE & ESD & RE & CP \\ 
		\midrule
		\multirow{5}{*}{ATT} 
		& Sand. & 0.03 &  1.56 & 1.26 & 1.56 & 0.55 & 0.99 & 0.15 & 1.61 & 2.17 & 1.61 & 0.55 & 0.99 \\ 			
		& WB R  & 0.03 & 1.56 & 0.92 & 1.56 & 1.06 & 0.94 & 0.15 & 1.61 & 2.01 & 1.61 & 0.64 &  0.98 \\  
		& WB E  & 0.03 & 1.56 & 0.91 & 1.56 & 1.12 & 0.93 & 0.15 & 1.61 &  1.95 & 1.61 & 0.68 & 0.98 \\
		& Std.Boot & 0.03 & 1.56 & 1.51 & 1.56 & 1.07 & 0.94 & 0.15 & 1.61 & 1.55 & 1.61 & 1.08 & 0.94 \\ 
		& TMLE & 0.75 & 1.71 & 1.58 & 1.70 & 1.16 & 0.94 & 0.70 & 1.74 & 2.07 & 1.74 & 0.71 & 0.98 \\ 
		\addlinespace
		\multirow{5}{*}{ATC}
		& Sand. & 0.01 & 0.68 & 0.88 & 0.68 & 0.59 & 0.99 & 6.33 & 5.71 &  3.38 & 5.62  & 2.78 &  0.88 \\  
		& WB R & 0.01 & 0.68 & 0.68 & 0.68 & 1.01 & 0.95 & 6.33  & 5.71 & 3.89 & 5.62  & 2.09 &  0.89 \\  
		& WB E & 0.01 & 0.68 & 0.67 & 0.68 & 1.03 & 0.94 & 6.33  & 5.71 & 3.25 & 5.62  & 2.99 & 0.85 \\
		& Std.Boot & 0.01 & 0.68 & 0.67 & 0.68 & 1.02 & 0.94 & 6.33 & 5.71 & 2.62 & 5.62 & 4.61 & 0.77 \\
		& TMLE & 12.68 & 4.57 & 3.37 & 4.08 & 1.47 & 0.79 & 7.06 & 3.88 & 2.98 & 3.71 & 1.55 & 0.89 \\ 
		\addlinespace 
		& &\multicolumn{6}{c}{PS and OR models misspecified} & \multicolumn{6}{c}{OR model correctly specified} \\\cmidrule(lr){3-8}\cmidrule(lr){9-14} 
		&  &  Bias & RMSE & SE & ESD & RE & CP &  Bias & RMSE & SE & ESD & RE & CP \\ 
		\midrule
		\multirow{5}{*}{ATT}  
		& Sand. & 2.19 & 1.94 & 2.38 & 1.89 & 0.63 & 0.97 & 0.04 & 1.56 & 2.10 & 1.56 & 0.55 & 0.99 \\ 
		& WB R  & 2.19 & 1.94 & 2.01 & 1.89 & 0.89 & 0.94 & 0.04 & 1.56 & 1.51 & 1.56 & 1.01 & 0.94 \\ 
		& WB E  & 2.19 & 1.94 & 1.96 & 1.89 & 0.92 & 0.94 & 0.04 & 1.56 & 1.47 & 1.56 & 1.12 & 0.93 \\ 
		& Std.Boot & 2.19 & 1.94 & 1.82 & 1.89 & 1.08 & 0.92 & 0.04 & 1.56 & 1.50 & 1.56 & 1.07 & 0.94 \\ 
		& TMLE & 1.48 & 2.04 & 2.03 & 2.02 & 0.99 & 0.94 & 1.43 & 1.63 & 1.49 & 1.60 & 1.15 & 0.91 \\ 
		\addlinespace 
		\multirow{5}{*}{ATC}
		& Sand. & 1.44 & 5.42 & 3.53 & 5.41 & 2.36 & 0.86 & 0.02 & 0.67 & 0.88 & 0.67 & 0.58 & 0.99 \\ 
		& WB R & 1.44 & 5.42 & 4.66 & 5.41 & 1.35  & 0.89 & 0.02 & 0.67 & 0.68 & 0.67  & 0.98 & 0.95 \\ 			  
		& WB E & 1.44 & 5.42 & 3.80 & 5.41 & 2.03 & 0.87 & 0.02 & 0.67 & 0.67 & 0.67  & 1.00 & 0.95 \\  
		& Std.Boot & 1.44 & 5.42 & 3.37 & 5.41 & 2.58 & 0.84 & 0.02 & 0.67 & 0.67 & 0.67 & 1.00 & 0.95 \\ 
		& TMLE & 2.87 & 3.27 & 2.95 & 3.23 & 1.20 & 0.86 & 24.95 & 5.03 & 3.09 & 2.97 & 0.93 & 0.82 \\ 
		\bottomrule
	\end{tabular}
	\begin{tablenotes}
		\tiny
		\item  Est.: Estimand; ATT (resp. ATC): average treatment effect on the treated (resp. controls); Bias: absolute relative bias$\times 100$;  
		\item  RMSE: root mean squared error; SE: median of standard errors from proposed method; ESD: empirical standard deviation; RE: median of relative efficiencies; CP: coverage probability; Sand.: sandwich; WB R (resp. E) : wild bootstrap via Rademacher (resp. exponential) distribution; Std.Boot: standarded bootstrap; TMLE: targeted maximum likelihood estimation.
	\end{tablenotes}
\end{threeparttable}
\end{table}	

\begin{table}\scriptsize 
\begin{threeparttable}
	\centering
	\caption{Model 2, with $p = 45.84\%$}
	\label{sim-model2.effect}
	\begin{tabular}{llcccccccccccc}
		\toprule	
		& & \multicolumn{12}{c}{Constant treatment effect} \\\cmidrule(lr){3-14}
		& &\multicolumn{6}{c}{PS and OR models correctly specified} & \multicolumn{6}{c}{PS model correctly specified} \\\cmidrule(lr){3-8}\cmidrule(lr){9-14} 
		Est. & Method & Bias & RMSE & SE & ESD & RE & CP & Bias & RMSE & SE & ESD & RE & CP \\ 
		\midrule
		\multirow{5}{*}{ATT} & Sand. &  0.04 & 0.16  & 0.24 & 0.16 & 0.43 & 1.00 & 4.31 & 1.16 &  0.90 & 1.14 & 1.63 &  0.93 \\ 			
		& WB R &  0.04 & 0.16 & 0.15 & 0.16 & 1.10 & 0.94 & 4.31 & 1.16 &  1.76 & 1.14 & 0.42 & 1.00 \\ 
		& WB E  &  0.04 & 0.16 & 0.15 & 0.16 & 1.12 &  0.94 & 4.31 & 1.16 &  1.66 & 1.14 & 0.47 & 1.00 \\
		& Std.Boot & 0.04 & 0.16 & 0.15 & 0.16 & 1.06 & 0.94 & 4.31 & 1.16 & 0.91 & 1.14 & 1.59 & 0.92 \\ 
		& TMLE & 0.04 & 0.16 & 0.15 & 0.16 & 1.11 & 0.94 & 5.66 & 1.04 & 1.50 & 1.02 & 0.46 & 0.99 \\ 
		\addlinespace
		\multirow{5}{*}{ATC}& Sand. &  0.05 & 0.17  & 0.24 & 0.17 & 0.52 & 0.99 & 9.57 & 2.84 & 2.21 & 2.81 & 1.62 &  0.96 \\  
		& WB R & 0.05 & 0.17 & 0.18 & 0.17 & 0.93 & 0.96 & 9.57 & 2.84 & 2.50 & 2.81 & 1.27 & 0.96 \\ 
		& WB E & 0.05 & 0.17 & 0.17 & 0.17 & 0.95 & 0.96 & 9.57 & 2.84 & 2.23 & 2.81 & 1.59 & 0.96 \\
		& Std.Boot & 0.05 & 0.17 & 0.16 & 0.17 & 1.06 & 0.94 & 9.57 & 2.84 & 1.47 & 2.81 & 3.68 & 0.82 \\ 
		& TMLE & 0.00 & 0.18 & 0.16 & 0.18 & 1.28 & 0.91 & 3.37 & 1.87 & 1.74 & 1.86 & 1.15 & 0.94 \\ 
		\addlinespace 
		& &\multicolumn{6}{c}{PS and OR models misspecified} & \multicolumn{6}{c}{OR model correctly specified} \\\cmidrule(lr){3-8}\cmidrule(lr){9-14} 
		&  & Bias & RMSE & SE & ESD & RE & CP & Bias & RMSE & SE & ESD & RE & CP \\ 
		\midrule
		\multirow{5}{*}{ATT}  & Sand. &52.45 & 2.85  & 1.69 & 1.93 & 1.30 & 0.82 & 0.03 & 0.15 &  0.24 & 0.15 & 0.41 & 1.00 \\ 
		& WB R  & 52.45 & 2.85 & 2.09 & 1.93 & 0.85 & 0.94 & 0.03 & 0.15 &  0.14 & 0.15 & 1.12 &  0.93 \\ 
		& WB E  & 52.45 & 2.85 &  1.93 & 1.93 & 0.97 & 0.91 & 0.03 & 0.15 & 0.14 & 0.15 & 1.14 &  0.93 \\ 
		& Std.Boot & 52.45 & 2.85 & 1.69 & 1.93 & 1.31 & 0.82 & 0.03 & 0.15 & 0.15 & 0.15 & 1.03 & 0.94 \\ 
		& TMLE & 49.41 & 2.63 & 1.66 & 1.73 & 1.09 & 0.86 & 0.03 & 0.15 & 0.15 & 0.15 & 1.10 & 0.94 \\ 
		\addlinespace 
		\multirow{5}{*}{ATC}& Sand. &  40.09 & 2.77  & 1.98 & 2.26 & 1.31 & 0.82 & 0.03 & 0.16 & 0.23 & 0.16 & 0.48 & 1.00 \\ 
		& WB R & 40.09 & 2.77 &  2.20 & 2.26 & 1.06 & 0.84 & 0.03 & 0.16 &  0.17 & 0.16 & 0.93 &  0.96 \\ 			  
		& WB E & 40.09 & 2.77 & 2.01 & 2.26 & 1.27 & 0.81 & 0.03 & 0.16 &  0.17 & 0.16 & 0.94 &  0.95 \\ 
		& Std.Boot & 40.09 & 2.77 & 1.82 & 2.26 & 1.54 & 0.77 & 0.03 & 0.16 & 0.16 & 0.16 & 1.02 & 0.94 \\ 
		& TMLE & 25.57 & 2.41 & 1.68 & 2.19 & 1.69 & 0.79 & 0.05 & 0.16 & 0.15 & 0.16 & 1.07 & 0.94 \\ 
		\addlinespace
		& & \multicolumn{12}{c}{Heterogeneous treatment effect} \\\cmidrule(lr){3-14}
		& &\multicolumn{6}{c}{PS and OR models correctly specified} & \multicolumn{6}{c}{PS model correctly specified} \\\cmidrule(lr){3-8}\cmidrule(lr){9-14} 
		Est. & Method & Bias & RMSE & SE & ESD & RE & CP & Bias & RMSE & SE & ESD & RE & CP \\ 
		\midrule
		\multirow{5}{*}{ATT} 
		& Sand. & 0.02 &  0.93 & 1.26 & 0.93 & 0.55 & 0.99 & 0.29 & 1.04 & 1.33 & 1.04 & 0.60 & 0.99 \\ 			
		& WB R  & 0.02 & 0.93 & 0.92 & 0.93 & 1.03 & 0.95 & 0.29 & 1.04 & 1.32 & 1.04 & 0.62 &  0.99 \\  
		& WB E  & 0.02 & 0.93 & 0.91 & 0.93 & 1.05 & 0.94 & 0.29 & 1.04 &  1.30 & 1.04 & 0.63 & 0.98 \\ 
		& Std.Boot & 0.02 & 0.93 & 0.92 & 0.93 & 1.03 & 0.95 & 0.29 & 1.04 & 1.01 & 1.04 & 1.06 & 0.95 \\ 
		& TMLE & 0.00 & 0.93 & 0.92 & 0.93 & 1.03 & 0.95 & 1.24 & 1.11 & 1.23 & 1.09 & 0.79 & 0.97 \\ 
		\addlinespace
		\multirow{5}{*}{ATC}
		& Sand. & 0.00 & 0.81 & 1.05 & 0.81 & 0.59 & 0.98 & 3.16 & 3.83 &  2.81 & 3.80  & 1.83 &  0.93 \\  
		& WB R & 0.00 & 0.81 & 0.85 & 0.81 & 0.90 & 0.95 & 3.16  & 3.83 & 3.19 & 3.80  & 1.42 &  0.93 \\  
		& WB E & 0.00 & 0.81 & 0.84 & 0.81 & 0.92 & 0.95 & 3.16  & 3.83 & 2.80 & 3.80  & 1.84 & 0.92 \\
		& Std.Boot & 0.00 & 0.81 & 0.78 & 0.81 & 1.06 & 0.94 & 3.16 & 3.83 & 1.98 & 3.80 & 3.68 & 0.80 \\ 
		& TMLE & 0.33 & 0.80 & 0.77 & 0.79 & 1.06 & 0.94 & 1.68 & 2.56 & 1.93 & 2.54 & 1.73 & 0.90 \\ 
		\addlinespace 
		& &\multicolumn{6}{c}{PS and OR models misspecified} & \multicolumn{6}{c}{OR model correctly specified} \\\cmidrule(lr){3-8}\cmidrule(lr){9-14} 
		&  & Bias & RMSE & SE & ESD & RE & CP & Bias & RMSE & SE & ESD & RE & CP \\ 
		\midrule
		\multirow{5}{*}{ATT}  & Sand. & 3.83 & 1.47 & 1.54 & 1.30 & 0.70 & 0.96 & 0.02 & 0.93 & 1.26 & 0.93 & 0.55 & 0.99 \\ 
		& WB R  & 3.83 & 1.47 & 1.38 & 1.30 & 0.89 & 0.93 & 0.02 & 0.93 & 0.92 & 0.93 & 1.03 & 0.95 \\ 
		& WB E  & 3.83 & 1.47 &1.35 & 1.30 & 0.92 & 0.93 & 0.02 & 0.93 & 0.91 & 0.93  & 1.06 & 0.95 \\ 
		& Std.Boot & 3.83 & 1.47 & 1.26 & 1.30 & 1.05 & 0.90 & 0.02 & 0.93 & 0.92 & 0.93 & 1.03 & 0.95 \\ 
		& TMLE & 4.49 & 1.57 & 1.23 & 1.33 & 1.18 & 0.87 & 0.21 & 0.92 & 0.90 & 0.92 & 1.03 & 0.95 \\ 
		\addlinespace 
		\multirow{5}{*}{ATC}& Sand. & 13.13 & 3.49 & 2.46 & 2.76 & 1.25 & 0.79 & 0.01 & 0.80 & 1.05 & 0.80 & 0.59 & 0.99 \\ 
		& WB R & 13.13 & 3.49 & 2.76 & 2.76 & 1.00  & 0.82 & 0.01 & 0.80 & 0.85 & 0.80  & 0.90 & 0.96 \\ 			  
		& WB E & 13.13 & 3.49 & 2.49 & 2.76 & 1.22 & 0.78 & 0.01 & 0.80 & 0.84 & 0.80  & 0.92 & 0.95 \\  
		& Std.Boot & 13.13 & 3.49 & 2.11 & 2.76 & 1.70 & 0.72 & 0.01 & 0.80 & 0.78 & 0.80 & 1.05 & 0.94 \\
		& TMLE & 7.91 & 2.84 & 1.91 & 2.54 & 1.77 & 0.80 & 0.10 & 0.82 & 0.79 & 0.82 & 1.07 & 0.94 \\ 
		\bottomrule
	\end{tabular}
	\begin{tablenotes}
		\tiny
		\item  Est.: Estimand; ATT (resp. ATC): average treatment effect on the treated (resp. controls); Bias: absolute relative bias$\times 100$;  
		\item  RMSE: root mean squared error; SE: median of standard errors from proposed method; ESD: empirical standard deviation; RE: median of relative efficiencies; CP: coverage probability; Sand.: sandwich; WB R (resp. E) : wild bootstrap via Rademacher (resp. exponential) distribution; Std.Boot: standarded bootstrap; TMLE: targeted maximum likelihood estimation.
	\end{tablenotes}
\end{threeparttable}
\end{table}	

\begin{table}\scriptsize 
\begin{threeparttable}
	\centering
	\caption{Model 3, with $p=79.22\%$}
	\label{sim-model3.effect}
	\begin{tabular}{llcccccccccccc}
		\toprule	
		& & \multicolumn{12}{c}{Constant treatment effect} \\\cmidrule(lr){3-14}
		& &\multicolumn{6}{c}{PS and OR models correctly specified} & \multicolumn{6}{c}{PS model correctly specified} \\\cmidrule(lr){3-8}\cmidrule(lr){9-14} 
		Est. & Method &  Bias & RMSE & SE & ESD & RE & CP &  Bias & RMSE & SE & ESD & RE & CP \\ 
		\midrule
		\multirow{5}{*}{ATT} 
		& Sand. & 0.21 & 0.22  & 0.25 & 0.22 & 0.81 & 0.98 & 9.89 & 2.48 &  1.51 & 2.45 & 2.63 & 0.86 \\ 			
		& WB R &  0.21 & 0.22 & 0.20 & 0.22 & 1.21 & 0.93 & 9.89 & 2.48 &  2.65 & 2.45 & 0.85 & 0.97 \\ 
		& WB E & 0.21 & 0.22 & 0.19 & 0.22 & 1.31 & 0.92 & 9.89 & 2.48 &  2.33 & 2.45 & 1.10 & 0.95 \\
		& Std.Boot & 0.21 & 0.22 & 0.21 & 0.22 & 1.07 & 0.95 & 9.89 & 2.48 & 1.72 & 2.45 & 2.02 & 0.90 \\ 
		& TMLE & 0.17 & 0.23 & 0.20 & 0.23 & 1.31 & 0.91 & 15.97 & 1.98 & 2.10 & 1.87 & 0.79 & 0.93 \\ 
		\addlinespace
		\multirow{5}{*}{ATC}
		& Sand. &  0.09 & 0.19  & 0.33 & 0.19 & 0.33 & 1.00 & 7.73 & 2.76 & 2.63 & 2.75 & 1.09 &  1.00 \\  
		& WB R & 0.09 & 0.19 & 0.28 & 0.19 & 0.44 & 1.00 & 7.73 & 2.76 & 2.90 & 2.75 & 0.90 & 1.00 \\ 
		& WB E & 0.09 & 0.19 & 0.28 & 0.19 & 0.45 & 1.00 & 7.73 & 2.76 & 2.70 & 2.75 & 1.03 & 1.00 \\
		& Std.Boot & 0.09 & 0.19 & 0.18 & 0.19 & 1.08 & 0.95 & 7.73 & 2.76 & 1.48 & 2.75 & 3.43 & 0.90 \\
		& TMLE & 0.00 & 0.19 & 0.18 & 0.19 & 1.15 & 0.94 & 16.64 & 1.91 & 2.35 & 1.79 & 0.58 & 1.00 \\
		\addlinespace 
		& &\multicolumn{6}{c}{PS and OR models misspecified} & \multicolumn{6}{c}{OR model correctly specified} \\\cmidrule(lr){3-8}\cmidrule(lr){9-14} 
		&  &  Bias & RMSE & SE & ESD & RE & CP &  Bias & RMSE & SE & ESD & RE & CP \\ 
		\midrule
		\multirow{5}{*}{ATT}  
		& Sand. & 112.98 & 6.23 & 3.01 & 4.29 & 2.03 & 0.80 & 0.16 & 0.22 & 0.25 & 0.22 & 0.78 & 0.98 \\ 
		& WB R & 112.98 & 6.23 & 4.35 & 4.29 & 0.97 & 0.98 & 0.16 & 0.22 &  0.21 & 0.22 & 1.12 &  0.94 \\ 
		& WB E & 112.98 & 6.23 & 3.69 & 4.29 & 1.35 & 0.94 & 0.16 & 0.22 & 0.20 & 0.22 & 1.21 &  0.93 \\ 
		& Std.Boot & 112.98 & 6.23 & 3.12 & 4.29 & 1.90 & 0.82 & 0.16 & 0.22 & 0.21 & 0.22 & 1.05 & 0.95 \\ 
		& TMLE & 85.15 & 4.60 & 2.90 & 3.09 & 1.14 & 0.91 & 0.04 & 0.21 & 0.21 & 0.21 & 1.02 & 0.95 \\ 
		\addlinespace 
		\multirow{5}{*}{ATC}
		& Sand. & 76.91 & 3.76 & 2.27 & 2.15 & 0.90 & 0.75 & 0.07 & 0.18 & 0.33 & 0.18 & 0.29 & 1.00 \\ 
		& WB R & 76.91 & 3.76 & 2.35 & 2.15 & 0.84 & 0.76 & 0.07 & 0.18 &  0.28 & 0.18 & 0.41 & 1.00 \\ 			  
		& WB E & 76.91 & 3.76 & 2.26 & 2.15 & 0.91 & 0.73 & 0.07 & 0.18 &  0.28 & 0.18 & 0.41 & 1.00 \\ 
		& Std.Boot & 76.91 & 3.76 & 1.96 & 2.15 & 1.21 & 0.63 & 0.07 & 0.18 & 0.18 & 0.18 & 0.99 & 0.95 \\ 
		& TMLE & 48.52 & 2.94 & 2.14 & 2.21 & 1.06 & 0.84 & 0.04 & 0.17 & 0.17 & 0.17 & 1.07 & 0.95 \\ 
		\addlinespace
		& & \multicolumn{12}{c}{Heterogeneous treatment effect} \\\cmidrule(lr){3-14}
		& &\multicolumn{6}{c}{PS and OR models correctly specified} & \multicolumn{6}{c}{PS model correctly specified} \\\cmidrule(lr){3-8}\cmidrule(lr){9-14} 
		Est. & Method &  Bias & RMSE & SE & ESD & RE & CP &  Bias & RMSE & SE & ESD & RE & CP \\ 
		\midrule
		\multirow{5}{*}{ATT} 
		& Sand. & 0.12 &  0.70 & 0.90 & 0.70 & 0.56 & 0.99 & 0.65 & 1.08 & 1.09 & 1.08 & 0.97 & 0.98 \\ 			
		& WB R  & 0.12 & 0.70 & 0.68 & 0.70 & 1.06 & 0.95 & 0.65 & 1.08 & 1.23 & 1.08 & 0.77 &  0.98 \\  
		& WB E  & 0.12 & 0.70 & 0.67 & 0.70 & 1.08 & 0.95 & 0.65 & 1.08 &  1.18 & 1.08 & 0.84 & 0.98 \\ 
		& Std.Boot & 0.12 & 0.70 & 0.68 & 0.70 & 1.05 & 0.95 & 0.65 & 1.08 & 0.94 & 1.08 & 1.30 & 0.94 \\ 
		& TMLE & 0.15 & 0.70 & 0.68 & 0.70 & 1.06 & 0.94 & 3.09 & 1.19 & 1.12 & 1.07 & 0.90 & 0.94 \\
		\addlinespace
		\multirow{5}{*}{ATC}
		& Sand. & 0.32 & 1.42 & 1.91 & 1.42 & 0.60 & 0.99 & 1.93 & 3.89 &  2.93 & 3.87  & 1.75 &  0.95 \\  
		& WB R & 0.32 & 1.42 & 1.75 & 1.42 & 0.66 & 0.98 & 1.93  & 3.89 & 3.36 & 3.87  & 1.33 &  0.96 \\  
		& WB E & 0.32 & 1.42 & 1.71 & 1.42 & 0.69 & 0.98 & 1.93  & 3.89 & 3.02 & 3.87  & 1.65 & 0.95 \\
		& Std.Boot & 0.32 & 1.42 & 1.40 & 1.42 & 1.03 & 0.94 & 1.93 & 3.89 & 2.10 & 3.87 & 3.42 & 0.84 \\
		& TMLE & 0.09 & 1.39 & 1.38 & 1.39 & 1.01 & 0.94 & 9.44 & 3.24 & 2.35 & 2.73 & 1.35 & 0.92 \\
		\addlinespace 
		& &\multicolumn{6}{c}{PS and OR models misspecified} & \multicolumn{6}{c}{OR model correctly specified} \\\cmidrule(lr){3-8}\cmidrule(lr){9-14} 
		&  &  Bias & RMSE & SE & ESD & RE & CP &  Bias & RMSE & SE & ESD & RE & CP \\ 
		\midrule
		\multirow{5}{*}{ATT}  
		& Sand. & 9.07 & 2.28 & 1.49 & 1.70 & 1.31 & 0.88 & 0.12 & 0.69 & 0.90 & 0.69 & 0.59 & 0.99 \\ 
		& WB R  & 9.07 & 2.28 & 1.65 & 1.70 & 1.06 & 0.95 & 0.12 & 0.69 & 0.68 & 0.69 & 1.04 & 0.95 \\ 
		& WB E  & 9.07 & 2.28 &1.50 & 1.70 & 1.28 & 0.92 & 0.12 & 0.69 & 0.67 & 0.69  & 1.06 & 0.94 \\ 
		& Std.Boot & 9.07 & 2.28 & 1.37 & 1.70 & 1.53 & 0.85 & 0.12 & 0.69 & 0.68 & 0.69 & 1.04 & 0.95 \\ 
		& TMLE & 6.16 & 1.77 & 1.30 & 1.44 & 1.23 & 0.90 & 0.48 & 0.68 & 0.68 & 0.67 & 0.97 & 0.96 \\ 
		\addlinespace 
		\multirow{5}{*}{ATC}
		& Sand. & 21.79 & 4.59 & 2.32 & 2.16 & 0.86 & 0.55 & 0.32 & 1.42 & 1.91 & 1.42 & 0.56 & 0.99 \\ 
		& WB R & 21.79 & 4.59 & 2.41 & 2.16 & 0.81  & 0.57 & 0.32 & 1.42 & 1.75 & 1.42  & 0.66 & 0.98 \\ 			  
		& WB E & 21.79 & 4.59 & 2.30 & 2.16 & 0.89 & 0.53 & 0.32 & 1.42 & 1.71 & 1.42  & 0.69 & 0.98 \\  
		& Std.Boot & 21.79 & 4.59 & 1.77 & 2.16 & 1.49 & 0.39 & 0.32 & 1.42 & 1.40 & 1.42 & 1.03 & 0.94 \\ 
		& TMLE & 10.82 & 2.70 & 1.94 & 1.80 & 0.87 & 0.78 & 1.13 & 1.58 & 1.47 & 1.57 & 1.14 & 0.94 \\ 
		\bottomrule
	\end{tabular}
	\begin{tablenotes}
		\tiny
		\item  Est.: Estimand; ATT (resp. ATC): average treatment effect on the treated (resp. controls); Bias: absolute relative bias$\times 100$;  
		\item  RMSE: root mean squared error; SE: median of standard errors from proposed method; ESD: empirical standard deviation; RE: median of relative efficiencies; CP: coverage probability; Sand.: sandwich; WB R (resp. E) : wild bootstrap via Rademacher (resp. exponential) distribution; Std.Boot: standarded bootstrap; TMLE: targeted maximum likelihood estimation.
	\end{tablenotes}
\end{threeparttable}
\end{table}	

\begin{table}\scriptsize 
\color{black}
\begin{threeparttable}
	\centering
	\caption{Model 4, with extreme weights and poor overlap}
	\label{sim-model4.effect}
	\begin{tabular}{llcccccccccccc}
		\toprule	
		& & \multicolumn{12}{c}{Constant treatment effect} \\\cmidrule(lr){3-14}
		& &\multicolumn{6}{c}{PS and OR models correctly specified} & \multicolumn{6}{c}{PS model correctly specified} \\\cmidrule(lr){3-8}\cmidrule(lr){9-14} 
		Est. & Method &  Bias & RMSE & SE & ESD & RE & CP &  Bias & RMSE & SE & ESD & RE & CP \\ 
		\midrule
		\multirow{5}{*}{ATT} 
		& Sand. & 0.17 & 0.37 & 0.34 & 0.37 & 1.22 & 0.95 & 21.67 & 3.10 & 1.55 & 2.98 & 3.68 & 0.78 \\ 
		& WB R & 0.17 & 0.37 & 0.25 & 0.37 & 2.23 & 0.88 & 21.67 & 3.10 & 1.98 & 2.98 & 2.26 & 0.89 \\ 
		& WB E & 0.17 & 0.37 & 0.23 & 0.37 & 2.75 & 0.85 & 21.67 & 3.10 & 1.82 & 2.98 & 2.66 & 0.87 \\ 
		& Std.Boot & 0.17 & 0.37 & 0.31 & 0.37 & 1.43 & 0.93 & 21.67 & 3.10 & 1.74 & 2.98 & 2.93 & 0.82 \\ 
		& TMLE & 0.20 & 0.39 & 0.23 & 0.39 & 2.99 & 0.77 & 14.32 & 3.86 & 2.00 & 3.82 & 3.65 & 0.76 \\  
		\addlinespace
		\multirow{5}{*}{ATC}
		& Sand.& 0.26 & 0.36 & 0.33 & 0.36 & 1.21 & 0.96 & 5.08 & 2.13 & 1.91 & 2.12 & 1.23 & 0.97 \\ 
		& WB R & 0.26 & 0.36 & 0.28 & 0.36 & 1.67 & 0.94 & 5.08 & 2.13 & 1.89 & 2.12 & 1.25 & 0.96 \\ 
		& WB E & 0.26 & 0.36 & 0.25 & 0.36 & 2.03 & 0.91 & 5.08 & 2.13 & 1.73 & 2.12 & 1.50 & 0.95 \\ 
		& Std.Boot & 0.26 & 0.36 & 0.28 & 0.36 & 1.61 & 0.92 & 5.08 & 2.13 & 1.66 & 2.12 & 1.63 & 0.94 \\ 
		& TMLE & 0.13 & 0.36 & 0.23 & 0.36 & 2.40 & 0.80 & 7.37 & 2.33 & 1.50 & 2.31 & 2.36 & 0.83 \\ 
		\addlinespace 
		& &\multicolumn{6}{c}{PS and OR models misspecified} & \multicolumn{6}{c}{OR model correctly specified} \\\cmidrule(lr){3-8}\cmidrule(lr){9-14} 
		&  &  Bias & RMSE & SE & ESD & RE & CP &  Bias & RMSE & SE & ESD & RE & CP \\ 
		\midrule
		\multirow{5}{*}{ATT}  
		& Sand. & 227.93 & 16.07 & 4.37 & 13.24 & 9.20 & 0.74 & 0.37 & 0.51 & 0.39 & 0.51 & 1.70 & 0.92 \\ 
		& WB R & 227.93 & 16.07 & 11.03 & 13.24 & 1.44 & 0.98 & 0.37 & 0.51 & 0.63 & 0.51 & 0.64 & 0.97 \\ 
		& WB E & 227.93 & 16.07 & 7.49 & 13.24 & 3.13 & 0.97 & 0.37 & 0.51 & 0.46 & 0.51 & 1.20 & 0.95 \\ 
		& Std.Boot & 227.93 & 16.07 & 4.83 & 13.24 & 7.52 & 0.82 & 0.37 & 0.51 & 0.38 & 0.51 & 1.77 & 0.94 \\ 
		& TMLE & 166.71 & 9.69 & 5.22 & 7.04 & 1.82 & 0.83 & 0.37 & 0.27 & 0.34 & 0.27 & 0.63 & 0.96 \\
		\addlinespace 
		\multirow{5}{*}{ATC}
		& Sand. & 136.89 & 7.29 & 3.06 & 4.81 & 2.47 & 0.73 & 0.07 & 0.30 & 0.31 & 0.30 & 0.90 & 0.96 \\ 
		& WB R & 136.89 & 7.29 & 3.61 & 4.81 & 1.77 & 0.89 & 0.07 & 0.30 & 0.25 & 0.30 & 1.38 & 0.92 \\ 
		& WB E & 136.89 & 7.29 & 2.95 & 4.81 & 2.65 & 0.78 & 0.07 & 0.30 & 0.23 & 0.30 & 1.61 & 0.91 \\ 
		& Std.Boot & 136.89 & 7.29 & 3.09 & 4.81 & 2.42 & 0.75 & 0.07 & 0.30 & 0.27 & 0.30 & 1.24 & 0.94 \\ 
		& TMLE & 178.98 & 8.82 & 2.57 & 5.16 & 4.03 & 0.42 & 0.00 & 0.28 & 0.22 & 0.28 & 1.58 & 0.89 \\  
		\addlinespace
		& & \multicolumn{12}{c}{Heterogeneous treatment effect} \\\cmidrule(lr){3-14}
		& &\multicolumn{6}{c}{PS and OR models correctly specified} & \multicolumn{6}{c}{PS model correctly specified} \\\cmidrule(lr){3-8}\cmidrule(lr){9-14} 
		Est. & Method &  Bias & RMSE & SE & ESD & RE & CP &  Bias & RMSE & SE & ESD & RE & CP \\ 
		\midrule
		\multirow{5}{*}{ATT} 
		& Sand. & 0.22 & 0.99 & 1.28 & 0.99 & 0.60 & 0.99 & 1.77 & 1.44 & 1.44 & 1.40 & 0.95 & 0.98 \\ 
		& WB R & 0.22 & 0.99 & 0.96 & 0.99 & 1.06 & 0.95 & 1.77 & 1.44 & 1.37 & 1.40 & 1.05 & 0.98 \\ 
		& WB E & 0.22 & 0.99 & 0.94 & 0.99 & 1.10 & 0.95 & 1.77 & 1.44 & 1.34 & 1.40 & 1.09 & 0.97 \\ 
		& Std.Boot & 0.22 & 0.99 & 0.96 & 0.99 & 1.05 & 0.95 & 1.77 & 1.44 & 1.18 & 1.40 & 1.40 & 0.94 \\ 
		& TMLE & 0.15 & 0.97 & 0.94 & 0.97 & 1.06 & 0.94 & 4.59 & 2.45 & 1.42 & 2.30 & 2.63 & 0.84 \\ 
		\addlinespace
		\multirow{5}{*}{ATC}
		& Sand. & 0.27 & 0.85 & 1.07 & 0.85 & 0.63 & 0.99 & 1.91 & 2.94 & 2.50 & 2.92 & 1.36 & 0.95 \\ 
		& WB R & 0.27 & 0.85 & 0.89 & 0.85 & 0.92 & 0.96 & 1.91 & 2.94 & 2.41 & 2.92 & 1.47 & 0.95 \\ 
		& WB E & 0.27 & 0.85 & 0.87 & 0.85 & 0.95 & 0.95 & 1.91 & 2.94 & 2.17 & 2.92 & 1.82 & 0.93 \\ 
		& Std.Boot & 0.27 & 0.85 & 0.81 & 0.85 & 1.09 & 0.94 & 1.91 & 2.94 & 2.30 & 2.92 & 1.62 & 0.94 \\ 
		& TMLE & 0.43 & 0.84 & 0.79 & 0.84 & 1.12 & 0.93 & 1.32 & 4.17 & 1.91 & 4.16 & 4.75 & 0.65 \\ 
		\addlinespace 
		& &\multicolumn{6}{c}{PS and OR models misspecified} & \multicolumn{6}{c}{OR model correctly specified} \\\cmidrule(lr){3-8}\cmidrule(lr){9-14} 
		&  &  Bias & RMSE & SE & ESD & RE & CP &  Bias & RMSE & SE & ESD & RE & CP \\ 
		\midrule
		\multirow{5}{*}{ATT}  
		& Sand. & 16.01 & 5.47 & 2.10 & 4.59 & 4.79 & 0.83 & 0.26 & 1.05 & 1.30 & 1.05 & 0.66 & 0.98 \\ 
		& WB R & 16.01 & 5.47 & 3.82 & 4.59 & 1.44 & 0.99 & 0.26 & 1.05 & 1.10 & 1.05 & 0.92 & 0.97 \\ 
		& WB E & 16.01 & 5.47 & 2.76 & 4.59 & 2.76 & 0.98 & 0.26 & 1.05 & 1.06 & 1.05 & 0.98 & 0.96 \\ 
		& Std.Boot & 16.01 & 5.47 & 2.05 & 4.59 & 5.00 & 0.86 & 0.26 & 1.05 & 0.99 & 1.05 & 1.12 & 0.94 \\ 
		& TMLE & 21.31 & 6.12 & 2.34 & 4.65 & 3.94 & 0.73 & 0.33 & 0.92 & 1.02 & 0.92 & 0.82 & 0.96 \\ 
		\addlinespace 
		\multirow{5}{*}{ATC}
		& Sand. & 45.92 & 9.63 & 4.04 & 6.33 & 2.45 & 0.75 & 0.27 & 0.83 & 1.06 & 0.83 & 0.60 & 0.99 \\ 
		& WB R & 45.92 & 9.63 & 4.80 & 6.33 & 1.74 & 0.90 & 0.27 & 0.83 & 0.87 & 0.83 & 0.90 & 0.96 \\ 
		& WB E & 45.92 & 9.63 & 3.87 & 6.33 & 2.67 & 0.78 & 0.27 & 0.83 & 0.86 & 0.83 & 0.92 & 0.95\\ 
		& Std.Boot & 45.92 & 9.63 & 4.01 & 6.33 & 2.49 & 0.74 & 0.27 & 0.83 & 0.81 & 0.83 & 1.05 & 0.94 \\ 
		& TMLE & 72.55 & 14.15 & 3.75 & 8.27 & 4.85 & 0.36 & 0.69 & 0.88 & 0.82 & 0.87 & 1.12 & 0.93 \\
		\bottomrule
	\end{tabular}
	\begin{tablenotes}
		\tiny
		\item  Est.: Estimand; ATT (resp. ATC): average treatment effect on the treated (resp. controls); Bias: absolute relative bias$\times 100$;  
		\item   RMSE: root mean squared error; SE: median of standard errors from proposed method; ESD: empirical standard deviation; RE: median of relative efficiencies; CP: coverage probability; Sand.: sandwich; WB R (resp. E) : wild bootstrap via Rademacher (resp. exponential) distribution; Std.Boot: standarded bootstrap; TMLE: targeted maximum likelihood estimation.
	\end{tablenotes}
\end{threeparttable}
\end{table}	

\begin{table}\scriptsize 
\color{black}
\begin{threeparttable}
	\centering
	\caption{Model 5a: Same as Model 1, but with $N=100$ and completed sandwich variance estimations}\label{sim-model5.effect}
	\begin{tabular}{llcccccccccccc}
		\toprule	
		& & \multicolumn{12}{c}{Constant treatment effect} \\\cmidrule(lr){3-14}
		& &\multicolumn{6}{c}{PS and OR models correctly specified} & \multicolumn{6}{c}{PS model correctly specified} \\\cmidrule(lr){3-8}\cmidrule(lr){9-14} 
		Est. & Method &  Bias & RMSE & SE & ESD & RE & CP &  Bias & RMSE & SE & ESD & RE & CP \\ 
		\midrule
		\multirow{5}{*}{ATT} 
		& Sand. & 0.00 & 0.64 & 1.06 & 0.64 & 0.36 & 0.99 & 28.27 & 3.52 & 3.05 & 3.34 & 1.20 & 1.00 \\ 
		& WB R & 0.00 & 0.64 & 0.57 & 0.64 & 1.27 & 0.93 & 28.27 & 3.52 & 5.30 & 3.34 & 0.40 & 1.00 \\ 
		& WB E & 0.00 & 0.64 & 0.53 & 0.64 & 1.47 & 0.90 & 28.27 & 3.52 & 4.63 & 3.34 & 0.52 & 1.00 \\ 
		& Std.Boot & 0.00 & 0.64 & 0.66 & 0.64 & 0.94 & 0.96 & 28.27 & 3.52 & 3.53 & 3.33 & 0.89 & 0.99 \\ 
		& TMLE & 10.93 & 0.90 & 0.59 & 0.78 & 1.74 & 0.85 & 16.98 & 3.75 & 4.89 & 3.69 & 0.57 & 0.99 \\  
		\addlinespace
		\multirow{5}{*}{ATC}
		& Sand. & 3.52 & 5.61 & 0.92 & 5.61 & 36.94 & 0.87 & 169.03 & 25.77 & 4.44 & 24.87 & 31.38 & 0.64 \\ 
		& WB R & 3.52 & 5.61 & 0.54 & 5.61 & 108.95 & 0.59 & 169.03 & 25.77 & 4.35 & 24.87 & 32.76 & 0.64 \\ 
		& WB E & 3.52 & 5.61 & 0.50 & 5.61 & 124.64 & 0.56 & 169.03 & 25.77 & 3.85 & 24.87 & 41.73 & 0.58 \\ 
		& Std.Boot & 3.52 & 5.61 & 14.99 & 5.61 & 0.14 & 1.00 & 169.03 & 25.78 & 9.48 & 24.88 & 6.89 & 0.91 \\ 
		& TMLE & 3.07 & 0.72 & 0.48 & 0.71 & 2.20 & 0.80 & 77.41 & 4.39 & 3.00 & 3.12 & 1.08 & 0.78 \\ 
		\addlinespace 
		& &\multicolumn{6}{c}{PS and OR models misspecified} & \multicolumn{6}{c}{OR model correctly specified} \\\cmidrule(lr){3-8}\cmidrule(lr){9-14} 
		&  &  Bias & RMSE & SE & ESD & RE & CP &  Bias & RMSE & SE & ESD & RE & CP \\ 
		\midrule
		\multirow{5}{*}{ATT}  
		& Sand. & 27.85 & 5.82 & 4.72 & 5.72 & 1.47 & 0.92  & 0.08 & 0.61 & 1.06 & 0.61 & 0.34 & 0.99 \\ 
		& WB R & 27.85 & 5.82 & 5.90 & 5.72 & 0.94 & 0.97 & 0.08 & 0.61 & 0.55 & 0.61 & 1.25 & 0.91 \\ 
		& WB E & 27.85 & 5.82 & 5.05 & 5.72 & 1.28 & 0.96 & 0.08 & 0.61 & 0.51 & 0.61 & 1.45 & 0.90 \\ 
		& Std.Boot & 27.85 & 5.82 & 5.00 & 5.72 & 1.30 & 0.95 & 0.08 & 0.61 & 0.63 & 0.61 & 0.96 & 0.95 \\ 
		& TMLE & 50.41 & 6.38 & 5.39 & 6.06 & 1.26 & 0.94 & 7.32 & 0.77 & 0.57 & 0.71 & 1.59 & 0.87 \\ 
		\addlinespace 
		\multirow{5}{*}{ATC}
		& Sand. & 149.81 & 26.03 & 4.42 & 25.34 & 32.87 & 0.61 & 3.64 & 5.61 & 0.93 & 5.61 & 36.71 & 0.87 \\ 
		& WB R & 149.81 & 26.03 & 4.42 & 25.34 & 32.83 & 0.63 & 3.64 & 5.61 & 0.53 & 5.61 & 110.08 & 0.59  \\ 
		& WB E & 149.81 & 26.03 & 3.93 & 25.34 & 41.50 & 0.57 & 3.64 & 5.61 & 0.51 & 5.61 & 123.00 & 0.56  \\ 
		& Std.Boot & 149.81 & 26.03 & 9.92 & 25.34 & 6.53 & 0.88 & 3.64 & 5.61 & 15.07 & 5.61 & 0.14 & 1.00 \\ 
		& TMLE & 73.01 & 5.42 & 3.00 & 4.57 & 2.32 & 0.68 & 2.51 & 0.68 & 0.51 & 0.68 & 1.79 & 0.84 \\ 
		\addlinespace
		& & \multicolumn{12}{c}{Heterogeneous treatment effect} \\\cmidrule(lr){3-14}
		& &\multicolumn{6}{c}{PS and OR models correctly specified} & \multicolumn{6}{c}{PS model correctly specified} \\\cmidrule(lr){3-8}\cmidrule(lr){9-14} 
		Est. & Method &  Bias & RMSE & SE & ESD & RE & CP &  Bias & RMSE & SE & ESD & RE & CP \\ 
		\midrule
		\multirow{5}{*}{ATT} 
		& Sand. & 0.49 & 4.97 & 6.24 & 4.97 & 0.64 & 0.96 & 1.28 & 5.34 & 6.65 & 5.33 & 0.64 & 0.96 \\ 
		& WB R & 0.49 & 4.97 & 4.49 & 4.97 & 1.23 & 0.90 & 1.28 & 5.34 & 5.94 & 5.33 & 0.81 & 0.94  \\ 
		& WB E & 0.49 & 4.97 & 3.89 & 4.97 & 1.64 & 0.86 & 1.28 & 5.34 & 5.12 & 5.33 & 1.08 & 0.92 \\ 
		& Std.Boot & 0.49 & 4.97 & 4.33 & 4.97 & 1.32 & 0.88 & 1.28 & 5.33 & 4.86 & 5.33 & 1.20 & 0.90 \\ 
		& TMLE & 8.49 & 5.49 & 4.29 & 5.19 & 1.46 & 0.91 & 7.68 & 6.33 & 5.54 & 6.12 & 1.22 & 0.93 \\ 
		\addlinespace
		\multirow{5}{*}{ATC}
		& Sand. & 1.47 & 6.94 & 2.85 & 6.94 & 5.92 & 0.96 & 55.47 & 34.26 & 5.52 & 33.06 & 35.85 & 0.61  \\ 
		& WB R & 1.47 & 6.94 & 2.11 & 6.94 & 10.85 & 0.86 & 55.47 & 34.26 & 5.33 & 33.06 & 38.45 & 0.60 \\ 
		& WB E & 1.47 & 6.94 & 1.97 & 6.94 & 12.43 & 0.84 & 55.47 & 34.26 & 4.70 & 33.06 & 49.54 & 0.54  \\ 
		& Std.Boot & 1.47 & 6.94 & 15.72 & 6.94 & 0.19 & 1.00 & 55.47 & 34.28 & 12.42 & 33.07 & 7.09 & 0.89 \\ 
		& TMLE & 7.15 & 2.55 & 1.79 & 2.27 & 1.60 & 0.80 & 30.51 & 6.96 & 3.20 & 4.88 & 2.33 & 0.54 \\  
		\addlinespace 
		& &\multicolumn{6}{c}{PS and OR models misspecified} & \multicolumn{6}{c}{OR model correctly specified} \\\cmidrule(lr){3-8}\cmidrule(lr){9-14} 
		&  &  Bias & RMSE & SE & ESD & RE & CP &  Bias & RMSE & SE & ESD & RE & CP \\ 
		\midrule
		\multirow{5}{*}{ATT}  
		& Sand. & 2.33 & 6.07 & 7.00 & 6.06 & 0.75 & 0.93 & 0.52 & 4.97 & 6.23 & 4.97 & 0.64 & 0.96 \\ 
		& WB R & 2.33 & 6.07 & 6.05 & 6.06 & 1.00 & 0.92 & 0.52 & 4.97 & 4.50 & 4.97 & 1.22 & 0.90 \\ 
		& WB E & 2.33 & 6.07 & 5.23 & 6.06 & 1.34 & 0.88 & 0.52 & 4.97 & 3.88 & 4.97 & 1.64 & 0.86 \\ 
		& Std.Boot & 2.33 & 6.07 & 5.34 & 6.06 & 1.28 & 0.88 & 0.52 & 4.97 & 4.31 & 4.97 & 1.33 & 0.89 \\ 
		& TMLE & 2.37 & 6.77 & 5.59 & 6.75 & 1.46 & 0.90  & 7.99 & 5.64 & 4.29 & 5.39 & 1.58 & 0.91 \\  
		\addlinespace 
		\multirow{5}{*}{ATC}
		& Sand. & 49.16 & 34.53 & 5.48 & 33.60 & 37.55 & 0.58 & 1.47 & 6.94 & 2.85 & 6.94 & 5.90 & 0.96 \\ 
		& WB R & 49.16 & 34.53 & 5.39 & 33.60 & 38.91 & 0.59 & 1.47 & 6.94 & 2.11 & 6.94 & 10.76 & 0.87  \\ 
		& WB E & 49.16 & 34.53 & 4.79 & 33.60 & 49.25 & 0.54  & 1.47 & 6.94 & 1.97 & 6.94 & 12.34 & 0.84 \\ 
		& Std.Boot & 49.16 & 34.53 & 13.00 & 33.60 & 6.68 & 0.87  & 1.47 & 6.94 & 14.86 & 6.94 & 0.22 & 1.00  \\ 
		& TMLE & 25.45 & 7.37 & 3.25 & 6.10 & 3.51 & 0.56  & 6.40 & 2.45 & 1.82 & 2.22 & 1.49 & 0.82 \\ 
		\bottomrule
	\end{tabular}
	\begin{tablenotes}
		\tiny
		\item  Est.: Estimand; ATT (resp. ATC): average treatment effect on the treated (resp. controls); Bias: absolute relative bias$\times 100$;  
		\item   RMSE: root mean squared error; SE: median of standard errors from proposed method; ESD: empirical standard deviation; RE: median of relative efficiencies; CP: coverage probability; Sand.: sandwich; WB R (resp. E) : wild bootstrap via Rademacher (resp. exponential) distribution; Std.Boot: standarded bootstrap; TMLE: targeted maximum likelihood estimation.
	\end{tablenotes}
\end{threeparttable}
\end{table}

\end{document}